# *An Index for SSRN Downloads*

Zura Kakushadze[§†1]

[§] *Quantigic® Solutions LLC,[2] 1127 High Ridge Road, #135, Stamford, CT 06905*

[†] *Free University of Tbilisi, Business School & School of Physics
240, David Agmashenebeli Alley, Tbilisi, 0159, Georgia*

September 5, 2015; revised November 13, 2015

*To my mother Ludmila (Mila) Kakushadze
on the occasion of her upcoming birthday*

Abstract

We propose a new index to quantify SSRN downloads. Unlike the SSRN downloads rank, which is based on the total number of an author's SSRN downloads, our index also reflects the author's productivity by taking into account the download numbers for the papers. Our index is inspired by – but is not the same as – Hirsch's $h$-index for citations, which cannot be directly applied to SSRN downloads. We analyze data for about 30,000 authors and 367,000 papers. We find a simple empirical formula for the SSRN author rank via a Gaussian function of the log of the number of downloads.

---

[1] Zura Kakushadze, Ph.D., is the President and a Co-Founder of Quantigic® Solutions LLC and a Full Professor in the Business School and the School of Physics at Free University of Tbilisi. Email: zura@quantigic.com

[2] DISCLAIMER: This address is used by the corresponding author for no purpose other than to indicate his professional affiliation as is customary in publications. In particular, the contents of this paper are not intended as an investment, legal, tax or any other such advice, and in no way represent views of Quantigic® Solutions LLC, the website www.quantigic.com or any of their other affiliates.

## 1. Introduction

In many scientific disciplines – e.g., physics – the total number of a researcher's citations is considered an important metric of the researcher's scientific impact. However, it does not take into account the author's productivity. An author may write just a single paper garnering many citations. Complementarily, the number of publications is also an important metric. However, it does not account for how important any of the researcher's (possibly numerous) papers are.

Hirsch (2005) proposed an index – the $h$-index[3] – whose purpose is to combine into a single number both the citations and publications figures. The appeal of the $h$-index is that, not only is it intuitive and simple to compute, it requires only data for a given author (publications and citations), no cross-sectional (across a sample of authors) data. E.g., INSPIRE High-Energy Physics Literature Database, Thomson Reuters Web of Science and Google Scholar all have utilized the $h$-index. Also, its variation has been applied to the internet media (Hovden, 2013).

Social Science Research Network (SSRN) keeps track of numbers of downloads for each author and paper. The number of SSRN downloads has – perhaps not surprisingly in hindsight – become an important metric in its own right. SSRN ranks authors and papers by the number of downloads. However, just as with citations, the number of downloads for a given author does not take into account the author's productivity, that is, the number of publications (or papers).

In this note we propose a new index for SSRN downloads. Unlike the SSRN downloads rank, which is based on the total number of an author's SSRN downloads, our index also reflects the author's productivity by taking into account the download numbers for the papers. Our index is inspired by – but is not the same as – the $h$-index. Thus, if we apply the $h$-index to SSRN downloads (via simply replacing citations by downloads), we will find that the $h$-index is mostly equal the number of papers: the bulk of the numbers of downloads exceeds the bulk of the numbers of citations by roughly a few orders of magnitude, so the $h$-index is uninformative.[4]

We circumvent this difficulty by noting that the numbers of downloads – in fact, just as the numbers of citations – have quasi-log-normal distributions. I.e., the numbers of downloads (citations) are exponential by nature.

---

[3] "A scientist has index $h$ if $h$ of his/her $N_p$ papers have at least $h$ citations each, and the other $(N_p - h)$ papers have no more than $h$ citations each" (Hirsch, 2005). This is the same as the Eddington number in cycling, with papers replaced by days and citations replaced by miles cycled on a given day. Its inventor, Sir Arthur Stanley Eddington (1882-1944), was a renowned British astronomer, physicist and mathematician and an avid cyclist.

[4] The $h$-index for citations turns out to be informative due to a numerological "accident" that the number of papers is typically bounded by (low) 2 digits, and the number of citations is bounded by roughly a square of that. The same does not hold for downloads, which roughly are a few orders of magnitude more ubiquitous. We could rescale the numbers of downloads by an overall factor based on cross-sectional data thereby forgoing simplicity.

We therefore define the following index – call it $k$ for the lack of a better name – for SSRN downloads for a given author:

$$k = \max_i \min(q_i, i) \tag{1}$$

$$q_i = \lfloor \ln(d_i) \rfloor \tag{2}$$

Where $d_i$ is the number of downloads for the $i$-th paper by said author; $i = 1, \ldots, n$; $n$ is the number of papers with $d_i > 0$; and $d_i$ are ordered decreasingly: $d_1 \geq d_2 \geq \cdots \geq d_n$. As we discuss in more detail below, the index $k$ appears to produce reasonable statistical output.

Let us recast (1) and (2) in plain English. For a given author, take all papers with nonzero downloads. Sort these $n$ papers decreasingly with the number of downloads. For each paper, take the integer part (the floor) of the natural log of the number of downloads, which is $q_i$. Let us call $q_i$ "log-downloads". Then the author's index equals $k$ if $k$ of the $n$ papers have at least $k$ log-downloads each, and the other $(n - k)$ papers have no more than $k$ log-downloads each. This is the same as the $h$-index with the numbers of citations replaced by log-downloads $q_i$.

The index $k$ is integer-valued. Currently, it varies from 0 to 9 (we discuss our data in detail in the next section). So, many authors share the same $k$. We can add granularity via non-integer indexes $k_*$ and $d_*$ (cf. (Ruane and Tol, 2008)):

$$k_* = \ln(d_*) = \frac{(k+1)\ln(d_k) - k\,s}{1 + \ln(d_k) - s} \tag{3}$$

Here $s = \ln(d_{k+1})$ if $k < n$; otherwise, $s = k$. By definition, $k \leq k_* < k+1$ and $k = \lfloor k_* \rfloor$. Eq. (3) has a simple geometric meaning. First, consider the case $k < n$. Once we fix $k$ via Eqs. (1) and (2), we linearly interpolate between the points $(k, \ln(d_k))$ and $(k+1, \ln(d_{k+1}))$ (cf. (Rousseau, 2006, 2014)). On that line there always exists a point $(k_*, k_*)$, which defines $k_*$. For $k = n$ we interpolate twixt $(k, \ln(d_k))$ and $(k+1, k)$; $k$ being a natural estimate for $\ln(d_{k+1})$. We compute $k$, $k_*$ and $d_*$ for about 30,000 SSRN authors (see Section 2). Table 1 gives the top 20 authors by their $k_*$ values and shows that the ranking of authors by $k_*$ does not coincide with their SSRN rank (similarly to ranking by the $h$-index v. ranking by citations). Figure 1 illustrates the computation of the $k$, $k_*$ and $d_*$ indexes. Both the $k_*$ and $d_*$ indexes are equally informative and as a matter of preference we can use either $k_*$ or $d_*$ interchangeably.

In Section 2 we discuss our data and its statistical characteristics for about 30,000 SSRN authors and over 367,000 papers. We find a simple empirical formula for the SSRN author rank $r$ (see Section 2 for the empirical values of the numeric coefficients $a, b, c$):

$$\ln(r) \approx a + b \ln(d_{tot}) + c \ln^2(d_{tot}) \tag{4}$$

where $d_{tot}$ is the author's total number of downloads. In section 3 we discuss some statistical properties of our indexes $k$, $k_*$ and $d_*$, including what is an analog of the ratio $h / n^2$ for the $h$-index. In Subsection 3.1 we discuss another index for SSRN downloads inspired by the $g$-index (Egghe, 2006) and how it compares to the $k_*$ index. We briefly conclude in Section 4, where we discuss advantages of our proposal, some caveats and (in some cases) how to cope with them.

As mentioned above, our indexes for SSRN downloads are related to the $h$-index (and, thereby, the Eddington number) by virtue of their definitions. In this regard, let us mention some prior works. Here we will not attempt a comprehensive overview of the literature on the $h$-index and various related indexes – for detailed reviews and extensive lists of references, see, e.g., (Alonso *et al*, 2009), (Egge, 2010) and (Norris and Oppenheim, 2010). Instead, here we focus on prior works with some potential relevance to the approach we follow in this paper.

Thus, various "nonlinear" generalizations of the $h$-index have been proposed, e.g., the $h(2)$ index (Kosmulski, 2006) and its generalizations (Levitt and Thelwall, 2007; Deineko and Woeginger, 2009). The $h(2)$ index was applied to article downloads as a metric for academic journals (Hua *et al*, 2009). Logarithms of citations have been considered in other contexts such as ranking (see, e.g., (Lundberg, 2007) and (Stringer, Sales-Pardo and Amaral, 2008)). However, to our knowledge, our log-based indexes, which stem from our observation of the exponential nature of SSRN downloads (and other metrics, including citations – see Subsection 3.5) are the first of their kind. Also, our application of indexes of this kind to SSRN downloads is novel and the beauty of working with SSRN downloads data is that it is large and provides lots of statistics. Currently, SSRN uses vanilla downloads (and, secondarily, citations) to rank authors and papers.

Variations on the $h$-index theme include the aforementioned $g$-index of (Egge, 2006) (which is analogous to the $h(2)$ index with citations replaced by cumulative citations), the $hg$-index (Alonso *et al*, 2010), the $h_\alpha$-index (Van Eck and Waltman, 2008), the $A$-index (Jin, 2006; Rousseau, 2006) and its variation the $m$-index (Bornmann, Mutz and Daniel, 2008), the $R$-index and the $AR$-index (Jin, 2007; Jin *et al*, 2007) (also see (Jarvelin and Persson 2008)), the citation-weighted $h$-index (Egge and Rousseau, 2008), the contemporary $h$-index, the trend $h$-index and the normalized $h$-index (Sidiropoulos, Katsaros and Manolopoulos, 2007), the dynamic $h$-type index (Rousseau and Ye, 2008), the tapered $h$-index (Anderson, Hankin and Killworth, 2008), variants accounting for multiple authors (Shreiber, 2008; Batista *et al*, 2006; Bornmann and Daniel, 2007; Imperial and Rodriguez-Navarro, 2007; Egge, 2008), and other variations of the $h$-index.

## 2. Data

In this section we describe our dataset. We downloaded all of our data directly from the SSRN website. The download was automated excepting some manual "patches" (see below).

SSRN provides the "Top Authors" data for the top 30,000 authors[5] based on downloads, both overall and for the last 12 months. We downloaded this data on 08/11/2015.[6] The data contains links to the authors' freely accessible individual webpages with their scholarly papers. Out of the 30,000 webpages 15 turned out to be "bad", so our dataset contains 29,985 authors.

The Top Authors data consists of 300 webpages, with 100 authors per page. Among other data, for each author these webpages contain the total number of downloads $d_{tot}$ and the total number of papers $n_{tot}$ (overall and for the last 12 months). Table 2 gives summaries for $d_{tot}$ and $n_{tot}$ as well as downloads-per-paper $d_{tot} / n_{tot}$ and their logarithms. Figure 2 plots densities and histograms for $\ln(d_{tot})$ (overall and 12 mo). The distribution for overall $\ln(d_{tot})$ is quasi-normal, so the distribution for overall $d_{tot}$ is quasi-log-normal, i.e., it is skewed with a long tail at the higher end. The distribution for 12-mo $d_{tot}$ is even more skewed. It is evident that we should not work with the numbers of downloads but their logs. This conclusion is further supported by the densities (and histograms) for downloads-per-paper in Figure 3.

Furthermore, the density for overall $\ln(d_{tot} / n_{tot})$ in Figure 3 is very close to a Gaussian. In Figure 4 we plot the same density together with a Gaussian curve from a least-squares fit.[7]

Here we should remark that the Top Authors data includes all papers in $n_{tot}$ and thereby in the $d_{tot} / n_{tot}$ computation, even those that SSRN does not include in the computation of $d_{tot}$ (such as an author's so-called "other papers").[8] Furthermore, there are regular papers with no downloads for "good reasons", e.g., papers with abstracts only and without downloadable PDF files. In many cases this is due to the policies of the journals where such papers are published, which do not permit posting published papers on the internet, including SSRN. Keeping such papers in $n_{tot}$ in the $d_{tot} / n_{tot}$ computation artificially lowers the downloads-per-paper figures. However, these nuances do not affect our conclusion relating to quasi-log-normality.

---

[5] SSRN provides the "Top Authors" data for all authors and also separately for the law, business and economics authors, with such data for the accounting and finance authors apparently forthcoming. We analyzed the data for all authors. It would be interesting to repeat our analysis for the above 5 disciplines (when they become available).

[6] Accessing this data beyond the top 10 authors requires an SSRN account login. The downloaded webpages state that the data was last updated on 07/27/2015.

[7] We use the R function `optim()` to determine the three parameters in the fit (mean, standard deviation and maximum value); see Figure 4.

[8] E.g., in the Top Authors data, P. Fernandez's (SSRN ID 12696) $d_{tot}$ is based on 230 of his papers. His 3 "other papers" (in the SSRN terminology) do not contribute to the total number of downloads; however, in the Top Authors data $n_{tot} = 233$ and this is the number used in computing $d_{tot} / n_{tot} = 808757 / 233 \approx 3471$. All figures are as of the date we downloaded the data (see above). Using $n_{tot} = 230$ would appear to be better.

*2.1. SSRN Rank v. Downloads*

The density plots in Figures 2 and 3 are rather convincing: the numbers of downloads are exponential by nature. Let $n_{auth}$ be the number of authors in our data. Then the SSRN author rank is given by (where the rank is computed across all authors in the Top Authors data)

$$r = n_{auth} + 1 - \text{rank}(d_{tot}) \qquad (5)$$

We plot $r$ v. $\ln(d_{tot})$ and $\ln(r)$ v. $\ln(d_{tot})$ in Figure 5 (overall and 12 mo). Let us start with the overall downloads. Except for the top 3 outliers (M.C. Jensen, P. Fernandez and E.F. Fama; see Table 1), the lower-left curve in Figure 5 is almost parabolic. A quadratic curve fits the data very well indeed. The results for the fit using a polynomial regression are given in Table 3. So, we have the empirical formula (4) with the numeric coefficients $a \approx 4.704$, $b \approx 2.009$, $c \approx -0.182$. Adding a cubic term does not improve the fit. The inflection point in the upper-left curve in Figure 5 occurs around $d_{tot} \approx 1300$. The results for the quadratic fit for the 12-mo downloads is summarized in Table 4. The empirical formula (4) also holds in this case with the numeric coefficients $a \approx 11.18$, $b \approx 0.492$, $c \approx -0.138$. A cubic term does not improve the fit. The inflection point in the upper-right curve in Figure 5 occurs around $d_{tot} \approx 40$.[9]

*2.2. Top Papers Data*

SSRN provides the "Top Papers" data for the top 10,000 papers[10] based on downloads, both overall and for the last 12 months. We downloaded this data on 08/13/2015.[11] The Top Papers data consists of 100 webpages, with 100 papers per page. Among other data, for each paper these webpages contain the number of downloads $d_p$ (overall and 12 mo). Table 5 gives summaries for $d_p$ and its log. Figure 6 plots densities and histograms for $\ln(d_p)$ (overall and 12 mo). The results are qualitatively similar to those for $\ln(d_{tot})$; see Table 2 and Figure 2.

Let $n_p$ be the number of papers in our data. Then the SSRN paper rank is given by (where the rank is computed across all papers in the Top Papers data)

$$r_p = n_p + 1 - \text{rank}(d_p) \qquad (6)$$

---

[9] There are more of what can be deemed as outliers in Figure 5, lower-right corner: P. Fernandez (139,290), M.C. Jensen (66,244), M.O. Jackson (41,580), M.T. Faber (37,553), A. Damodaran (34,523), C.R. Harvey (32,171), H.M. Mialon (32,072), E.F. Fama (32,060), C.R. Sunstein (31,681), B. Bartlett (31,583) and A.M. Francis (31,517), with the total number of downloads $d_{tot}$ in the last 12 months given in the parentheses.

[10] Unlike the Top Authors data, the Top Papers data does not appear to be available by (any) discipline.

[11] Accessing this data beyond the top 10 papers requires an SSRN account login. The downloaded webpages state that the data was last updated on 08/09/2015.

We plot $r_p$ v. $\ln(d_p)$ and $\ln(r_p)$ v. $\ln(d_p)$ in Figure 7 (overall and 12 mo). Let us start with the overall downloads. As above, except for several top outliers,[12] the lower-left curve in Figure 7 is almost parabolic. A quadratic curve fits the data very well indeed. The results for the fit using a polynomial regression are given in Table 6. So, we have the empirical formula (4) (where $r$ is replaced by $r_p$ and $d_{tot}$ is replaced by $d_p$) with the numeric coefficients $a \approx 4.375$, $b \approx 1.952$, $c \approx -0.199$.[13] The inflection point in the upper-left curve in Figure 7 occurs around $d_p \approx 650$. The results for the quadratic fit for the 12-months downloads are summarized in Table 7, so we have $a \approx 16.78$, $b \approx -1.331$, $c \approx -0.031$ in the formula (4).[14] However, the curvature is negligible, so we can use a linear fit instead by setting $c = 0$ in Eq. (4), for which the results are provided in Table 8, and we have $\ln(r_p) \approx a + b\ln(d_p)$ with $a \approx 17.95$, $b \approx -1.715$.

## *2.3. Data from SSRN Author Pages*

As mentioned above, the Top Authors data contains links to the 30,000 authors' individual webpages. We downloaded these webpages in an automated fashion on 08/16/2015 and 08/17/2015.[15] The data is essentially structured, with a few caveats. E.g., the "Posted:" date is not always shown, which complicates parsing. Also, the default ordering of the papers is by the decreasing number of downloads; however, occasionally this ordering is not followed with no clear pattern. Furthermore, papers that have been revised and are still under review by SSRN are moved to the bottom of the list with no "Last Revised:" date. However, each webpage has a field showing the total number of downloads, so a simple "sanity check" is that summing the number of paper downloads over all papers with non-zero/non-empty downloads fields should produce the total number of downloads. Out of 29,985 good webpages (see above), all but 256 satisfied this criterion with straightforward parsing. An additional heuristic further reduced this number to 78. We manually checked and "patched" the data for these remaining 78 pages on 08/18/2015. However, the laborious and time-consuming sourcing resulted in high quality data.

---

[12] The following papers are the apparent outliers (the format is "(authors(s), year), $d_p$"): (Faber, 2007), 152,242; (Solove, 2007), 150,263; (Jensen and Meckling, 1976), 108,410; (Jackson, 2011), 96,296; (Fama, 1998), 83,174; (Girgis, George and Anderson, 2010), 73,622. We include these papers in the References so the reader can get a flavor on the spectrum of the topics and authors of the top downloaded papers (overall).

[13] The coefficients $b$ and $c$ in this case are close to those for the overall total downloads (see above and Table 3).

[14] The following papers are the apparent outliers (the format is "(authors(s), year), $d_p$"): (Francis and Mialon, 2014), 31862 ; (Jackson, 2011), 31476; (Bartlett, 2015), 30538; (Faber, 2007), 21531; (Fama and French, 2015), 12081; (Roche, 2011), 12078. Again, we include these papers in the References so the reader can get a flavor on the spectrum of the topics and authors of the top downloaded papers (12 mo).

[15] The downloads take some time, even for a fast machine with a fast internet connection, which is what we used. We mention this to emphasize that the data is not 100% "synchronized" as, e.g., SSRN updates download counts in "real-time". This "asynchronicity" is unavoidable with downloads; however, its effect on our analysis here is small.

For the reasons mentioned in Section 2, we drop all papers labeled as "other papers" (SSRN does not include downloads for such papers in the total download count), and also all papers with empty downloads fields. For each author we then have a vector $d_i$, $i = 1, \ldots, n_{tot}$ with $\sum_{i=1}^{n_{tot}} d_i = d_{tot}$, same as the total number of downloads on the author's webpage. However, $n_{tot}$ can be less than the total number of papers on the webpage as we omit the papers with empty downloads fields. Using this data we compute $k$, $k_*$ and $d_*$ via Eqs. (1), (2) and (3).

## 3. Index Properties

The number of papers (as defined above) across all authors in our database is 367,478. However, by definition, only a fraction of these papers contribute to the index $k$: the number of such papers is simply a sum (across all authors) over the values of the integer index $k$ and turns out to be 112,793, or about 30.7%. Table 9 gives cross-sectional (across all authors in the SSRN Top Authors data) summaries for $k_*$ and the ratio $u = k / n_{tot}$ (with NAs omitted). The $u = 1$ cases are rather ubiquitous, to wit, 8,572. However, these are mostly the authors with low paper counts. We have the following statistics for the number of occurrences of $k = n_{tot}$ according to the $n_{tot}$ value: 3,326 for $n_{tot} = 1$; 2,338 for $n_{tot} = 2$; 1,842 for $n_{tot} = 3$; 848 for $n_{tot} = 4$; 205 for $n_{tot} = 5$; 12 for $n_{tot} = 6$; 0 for $n_{tot} = 7$; 1 for $n_{tot} = 8$; and 0 for $n_{tot} = 9$ (see the histogram in Figure 8). The outlier $k = n_{tot} = 8$ corresponds to the author G. Feiger (SSRN ID 1325770), whose $d_i = (47043, 7971, 7205, 6438, 6004, 5607, 5397, 5276, 3145)$.

The index $k_*$ means that $k = \lfloor k_* \rfloor$ papers have at least $d_k$ downloads each, i.e., we have $d_{tot} \geq k\, d_k$ and we can define the ratio

$$w = \frac{d_{tot}}{k\, d_k} \tag{7}$$

We have $w \geq 1$. Summaries of $w$ and $\ln(w)$ are given in Table 9. The large values of $w$ are mostly due to the authors with low paper counts but substantial numbers of downloads. In Figure 8 we plot histograms for $\ln(w)$, $k_*$ and the quantity $v = \ln(d_{tot} / n_{tot}) / n_{tot}$ also summarized in Table 9. The significance of the quantity $v$ is that, if all $q_i \geq n_{tot}$ (see Eq. (2)), then $k = n_{tot}$ and $\ln(w) \leq v$. The reason why the $k_*$ index works "numerologically", that is, produces reasonable results, is that the bulk of the values of $v$ are of order 1. Had these values been much higher, most values of $k_*$ would equal $n_{tot}$ and this index would be uninformative. The analog of the quantity $v$ for the $h$-index is $d_{tot} / n_{tot}^2$ and this quantity is mostly large for SSRN downloads, which is precisely why the $h$-index does not work "numerologically" for SSRN downloads. In contrast, for citations the bulk of the values of $c_{tot} / n_{tot}^2$ (here $c_{tot}$ is the total number of citations) is around 3-5 for physics papers (Hirsch, 2005) focuses on, which is the reason the $h$-index works reasonably well "numerologically" for citations in that particular field.

### *3.1. An Alternative Index*

Suppose an author has index $k_*$. This index knows nothing about the detailed structure of the downloads for the papers with $i \leq k = \lfloor k_* \rfloor$, only that $d_i \geq d_* = \exp(k_*)$ (we are assuming that the papers are ordered with decreasing $d_i$). To give more weight to the papers with more downloads, we can consider an alternative index – call it $\kappa$ for the lack of a better name – for SSRN downloads for a given author:

$$\kappa = \max_i \min(t_i, i) \tag{8}$$

$$t_i = \lfloor \ln(f_i) \rfloor \tag{9}$$

$$f_i = \frac{1}{i} \sum_{j=1}^{i} d_j \tag{10}$$

I.e., the integer-valued index $\kappa$ is based on the average number of downloads for the first $\kappa$ papers (as opposed to the number of downloads for the $\kappa$-th paper).[16] As in Section 1, we can define a non-integer index $\kappa_*$ via (see Figure 9)

$$\kappa_* = \ln(f_*) = \frac{(\kappa + 1)\ln(f_\kappa) - \kappa\, \sigma}{1 + \ln(f_\kappa) - \sigma} \tag{11}$$

Here $\sigma = \ln(f_{\kappa+1})$ if $\kappa < n_{tot}$; otherwise, $\sigma = \kappa$. By construction, $\kappa \leq \kappa_* < \kappa + 1$ and $\kappa = \lfloor \kappa_* \rfloor$. The analog of $w$ in Eq. (7) is $\omega = d_{tot} / \kappa\, f_\kappa$; however, the analog of $v$ is trivial: for $\kappa = n_{tot}$ we have $\omega = 1$ as $f_\kappa = d_{tot} / n_{tot}$ in this case. By construction, the number of papers that contribute to the index $\kappa$ is higher (compared with the index $k$); it is simply a sum (across all authors) over the values of the integer index $\kappa$ and turns out to be 128,494, or about 35.0%. Table 10 gives cross-sectional (across all authors in the SSRN Top Authors data) summaries for $\omega$, $\ln(\omega)$, $\kappa_*$ and the ratio $\mu = \kappa / n_{tot}$ (with NAs omitted). As above, the $\mu = 1$ cases are rather ubiquitous, to wit, 11,343, and mostly correspond to the authors with low paper counts. We have the following statistics for the number of occurrences of $\kappa = n_{tot}$ according to the $n_{tot}$ value (currently, the maximum value of $\kappa$ is 10): 3,326 for $n_{tot} = 1$; 2,395 for $n_{tot} = 2$; 2,206 for $n_{tot} = 3$; 1,868 for $n_{tot} = 4$; 1,141 for $n_{tot} = 5$; 356 for $n_{tot} = 6$; 42 for $n_{tot} = 7$; 8 for $n_{tot} = 8$; 1 for $n_{tot} = 9$; and 0 for $n_{tot} = 10$ (see the histogram in Figure 10). The outlier $\kappa = n_{tot} = 9$ corresponds to the author S. Zafron (SSRN ID 279069), whose 9 papers have the downloads vector $d_i = (21609, 21148, 17194, 11034, 8930, 1308, 1134, 256, 231)$. Table 11 lists the top 20 authors by the index $\kappa_*$ (cf. Table 1 for the top 20 authors by the index $k_*$).

---

[16] This is analogous to the $g$-index (Egghe, 2006) for citations. Here we have (the integer part of) the log in Eq. (9), which makes all the difference. Just as the $h$-index, the $g$-index is uninformative when applied to SSRN downloads.

*3.2. Are Our Indexes Informative?*

One of the critiques of the $h$-index was set forth in (Yong, 2014). In a nutshell, it boils down to the fact that we can think of $c_{tot}$ citations being partitioned into $n_{tot}$ papers, and then the $h$-index is the side-length of the so-called Durfee square (which is the largest $h \times h$ square that fits into the so-called Young diagram for said partition (see, e.g., (Anderson, Hankin and Killworth, 2008))). For given $c_{tot}$ and $n_{tot}$ there is a finite range of what values the $h$-index can take. Assuming equal probabilities (i.e., no additional information) we can define the expected value of the $h$-index. When $c_{tot}$ is large, there is an asymptotic formula for this expected value (Canfield, Corteel and Savage, 1998), which Yong (2014) proposes to use as the "rule-of-thumb" estimate for the $h$-index: $h = (\sqrt{6} \ln(2) / \pi) \sqrt{c_{tot}} \approx 0.54 \sqrt{c_{tot}}$. Yong then argues that the information in the $h$-index beyond what is already in $c_{tot}$ is limited. In this regard, here we also should ask whether our indexes $k_*$ and $\kappa_*$ are informative (beyond what is encoded in $d_{tot}$).

Since our indexes are based on logs of the numbers of downloads, the combinatorial tricks do not appear to be directly applicable. We will therefore take an empirical approach. We plot the indexes $k_*$ and $k$ as well as $\kappa_*$ and $\kappa$ v. $\ln(d_{tot})$ in Figure 11. There is an apparent linear $\ln(d_{tot})$ component in these indexes. Linear regressions of the indexes over $\ln(d_{tot})$ with the intercept are summarized in Tables 12-15. Adding another explanatory variable $\ln(n_{tot})$ improves the fits – see Table 16. Evidently, the dependence on $\ln(d_{tot})$ is not the end of the story: there is more information encoded in these indexes beyond what is already in $\ln(d_{tot})$.

*3.3. Twelve-months Indexes*

One "shortcoming" of the $h$-index is that, for a given author, it cannot decrease. A retired author can have a high $h$-value without writing a single new paper. By definition, the same applies to our indexes. In the case of the $h$-index, one can implement a weighting scheme, whereby older papers are given less weight (Sidiropoulos, Katsaros and Manolopoulos, 2007). The same idea can be applied to our indexes. However, the data for the age of the downloads is not readily available, at least not publically, so any empirical analysis presently is out of reach.

Nonetheless, not all is lost. We can compute our indexes for the last 12-months downloads. The author webpages do not separately provide the last 12-mo download data. We circumvent this difficulty by utilizing the Top Papers data, which contains the top 10,000 most downloaded papers for the last 12 months together with their last 12-mo download numbers.[17] For each author contributing to these 10,000[18] papers we can extract the author's papers with the last 12-mo download numbers and therefore compute our indexes. Summaries for the 12-mo $k_*$ and $\kappa_*$ index values are given in Table 17. Tables 18 and 19 give the top 20 authors by the 12-mo $k_*$ and $\kappa_*$ index values, respectively. The numbers of papers (column 6) in Tables 18 and 19 are lower than those in Table 1. It is not surprising that some papers do not make it to the top 10,000 most downloaded papers. This causes the total numbers of downloads in the last 12 months in Tables 18 and 19 to be lower than those reported in the Top Authors data (not shown), so the 12-mo rank in Tables 18 and 19 (column 7), which is based on the total numbers of downloads in Tables 18 and 19 (column 5), is not always the same as the 12-mo rank in the Top Authors data. However, we expect that the omitted low-downloads papers hardly affect the index values. The rank is not critical here and is shown solely for comparison purposes.

---

[17] It also contains data for the overall downloads. However, the Top Papers data is ordered by the rank based on the last 12-mo downloads. This implies that this data may not contain the overall top 10,000 most downloaded papers. The same applies to the Top Authors data, which is also ordered by the rank based on the last 12-mo downloads. However, it is not unreasonable to focus on the papers that have been downloaded more recently.

We can analyze the time-dependence of our indexes in more detail using the SSRN author webpage data. It contains the "Posted:" field (with some missing cases – see above), the date a paper was originally posted on SSRN. We parsed the 30,000 author webpages (see above) and for each author identified the earliest of the "Posted:" dates, which we use to measure $T$, the authors' "SSRN career" lengths in years (for simplicity we set 1 mo = 1/12 yr, 1 day = 1/30 mo).

Plots of $\ln(d_{tot})$, $\ln(d_{tot}/n_{tot})$, $k_*$ and $\kappa_*$ v. $\ln(T)$ are given in Figure 12.[19] There is no statistically significant relation between $\ln(d_{tot}/n_{tot})$ and $\ln(T)$; see Table 20.[20] On average, there is linear growth in $\ln(d_{tot})$, $k_*$ and $\kappa_*$ (at the upper end of the values) with $\ln(T)$, which is expected; see Table 20. Adding $\ln(T)$ as a third explanatory variable in the regressions in Table 16, however, has a negligible effect on the fits. The $\ln(T)$ dependence is not the main driver.[21]

### *3.4. Why Do Our Indexes Work "Numerologically"?*

In our definition of the index $k$ in Eqs. (1) and (2) (and, consequently, in our definitions of the indexes $k_*$ as well as $\kappa$ and $\kappa_*$) we chose the natural logarithm $\ln(\cdot)$ as opposed to a logarithm $\log_\beta(\cdot)$ with another base $\beta$. How come? The answer is rather prosaic. We chose the natural logarithm because it works well "numerologically". Let us elaborate on this point.

---

[18] One paper does not list the author(s) in Top Papers, so we end up with 9,999 papers with 11,871 authors.

[19] We downloaded the 30,000 SSRN author webpages to extract the "Posted:" fields on August 28-29, 2015. This does not affect the actual values of $T$; however; there were more "bad" webpages, 27 instead of 15 (see above).

[20] So that the statistic is meaningful, we take $T \geq 1$. After dropping NAs (see above), we have 28,552 datapoints. The summary for $T$ reads: Min = 1.003, 1st Quartile = 4.111; Median = 7.467; Mean = 8.415; 3rd Quartile = 12.060; Max = 21.270. The maximum corresponds to the author J. Pontiff (SSRN ID 17153), with a posting on 5/9/1994.

[21] Cf. the so-called $m$-quotient (the $h$-index over the number of years); see (Hirsch, 2005).

Suppose we have $n$ objects (e.g., papers), each of which is characterized by a count of sorts (e.g., a number of citations or downloads, etc.). Let us call these counts $z_i$, $i = 1, \ldots, n$. Let us further assume that the counts are exponential by nature (just as is the case with downloads), i.e., the cross-sectional distribution of the total counts $z_{tot} = \sum_{i=1}^{n} z_i$ across the object owners (e.g., authors) is (quasi-)log-normal. Suppose we wish to construct an index along the lines of our $k$ index. We can define this index via Eq. (1) with $q_i$ defined more generally as follows:[22]

$$q_i = \lfloor \gamma \ln(z_i) \rfloor \qquad (12)$$

Here $\gamma$ is an overall normalization factor, which for SSRN downloads in Eq. (2) we have set to 1 for the reasons we will explain momentarily. More generally, $\gamma$ need not be 1. The choice of the base in the logarithm is then subsumed in $\gamma$ as $\log_\beta(\cdot) = \ln(\cdot) / \ln(\beta)$. I.e., the choice of the base $\beta$ of the logarithm is equivalent to the choice of the overall normalization factor $\gamma$. In the context of the $h$-index this is analogous to the $h_\alpha$-index of (Van Eck and Waltman, 2008); also see (Waltman and Van Eck, 2009). Our $\gamma$ in Eq. (12) is analogous to $\alpha$ in the $h_\alpha$-index.

So, what should we choose as our factor $\gamma$? There is no magic prescription here. There are two evident guiding principles: i) that the resulting index $k$ (and also $k_*$ and all the other related indexes) should be informative, and ii) simplicity. E.g., if we choose $\gamma$ too high, the index $k$ will mostly equal the number of objects $n$ and thereby be uninformative. If we choose $\gamma$ too low, then the index $k$ will mostly equal 0 or 1 and thereby also be uninformative. A choice of $\gamma$ that avoids such extremes is such that the bulk of the values of the product $\gamma\,v$ is of order 1, where, as above, $v = \ln(z_{tot} / n) / n$. E.g., we can set $\gamma = 1 / \text{median}(v)$, albeit this is not the only choice. For the overall SSRN downloads the bulk of the values of $v$ is of order 1 (see Table 9), which is why we have chosen $\gamma = 1$ in Eq. (2), or, equivalently, the natural logarithm and not any other base. As mentioned above, this is the "numerological" reason why our indexes work well for the overall SSRN downloads. Also, while, e.g., $\text{median}(v) \approx 0.7$ (see Table 9), we have chosen $\gamma = 1$ based on a further consideration of simplicity, so that only each author's data is required to compute his/her indexes (but no cross-sectional data across a sample of authors).

Here the following remark is in order. Basing $\gamma$ on the quantity $v = \ln(z_{tot} / n) / n$ makes sense only if $\ln(z_{tot} / n)$ is essentially normally distributed and the data is not inundated with $n = 1$ (and, more generally, low $n$) datapoints. Our dataset based on the Top Authors data satisfies these criteria: as we discussed above, the density of $\ln(d_{tot} / n_{tot})$ for the overall downloads is almost Gaussian (see Figure 4), among the 29,979 datapoints (there are 29,985 "good" webpages (see above), 6 of which contain no papers) there are only 3,326 papers with $n_{tot} = 1$ and 2,399 papers with $n_{tot} = 2$, and the paper count statistics is reasonable (see

---

[22] Let us note that rescaling $z_i \rightarrow \theta\, z_i$ by some factor $\theta$ would merely shift the range of values of the index $k$.

Table 2). If $\ln(z_{tot} / n)$ itself has a highly skewed distribution or the data mostly contains low $n$ points, then blindly relying on $v = \ln(z_{tot} / n) / n$ would produce nonsensical results. E.g., median$(v) \approx 5.3$ for the 9,999 papers discussed in Subsection 3.3 based on the Top Papers data, despite the fact that the bulk of the 12-mo download numbers are roughly 5-7 times less numerous than the bulk of the overall download numbers. This is due to the fact that the majority of the 11,871 authors of these 9,999 papers have $n_{tot} = 1$ and $n_{tot} = 2$ (see Table 17). If we remove the $n_{tot} = 1$ and $n_{tot} = 2$ datapoints, happily we are left with only 1,310 authors with the bulk of the values of $v$ of order 1 (to wit, median$(v) \approx 1.4$).

### 3.5. Can We Apply Our Indexes to Citations?

As mentioned above, even citations are essentially exponential by nature. In this regard, we believe it would make sense to apply our ideas here to citations as well. This interesting in its own right topic is outside of the scope of this paper, so we will not delve into it too deeply and only give a bird's-eye view. A detailed empirical analysis would be required to see if it works.

If we apply, say, the $k$ index as defined in Eqs. (1) and (2) directly to citations, it may not work as well "numerologically". This is because the numbers of citations are a few orders of magnitude lower than the numbers of SSRN downloads. Thus, as of 9/4/2015, M.C. Jensen has the most SSRN downloads, to wit, 830,936, while according to INSPIRE (see above) E. Witten's (high energy physics) total number of citations is 118,374 with a total of 332 citable papers.[23] As of 9/4/2015, (Faber, 2007) is the most downloaded SSRN paper with 158,804 downloads, while the most cited paper in high energy physics (per INSPIRE) is (Maldacena, 1997) with 10,996 citations. As is customary in high energy physics, we exclude (Particle Data Group Collaboration, 2014) with 50,004 citations as of the end of 2014, which is a "handbook" of elementary particles and traditionally garners most citations in high energy physics year after year. Based on the above numbers, we can superficially estimate that there is roughly 1-1.5 orders of magnitude difference between SSRN downloads and citations, albeit a more detailed analysis (which is outside of the scope of this paper) would be required to get more precise bulk numbers. In any event, if we apply our indexes with $\gamma = 1$ to citations, we can expect that they will produce reasonable results at the higher end (i.e., for highly cited authors and papers), and a nontrivial $\gamma$ might be required to have informative indexes for lower citation count trenches.

Our log-based indexes should be applicable beyond SSRN downloads and citations, e.g., for internet media downloads, which apparently are also exponential by nature (cf. (Hovden, 2013)). However, depending on a type of downloads and the bulk values thereof, the factor $\gamma$ in Eq. (12) may have to be chosen away from 1 for our indexes to work well "numerologically".

---

[23] Using Harzing's Publish or Perish software (version 4.19.0.5725) with Google Scholar as the data source gives inflated figures: 161,969 citations and 574 papers. In our experience, INSPIRE does undercount citations, though.

## 4. Conclusions

Let us start by tying up a "loose end", so to speak. In the empirical regression in Table 6 we used the Top Papers data, which only contains 10,000 datapoints. We did so for illustrative purposes as downloading this data, which amounts to downloading only 100 webpages, is much less arduous than downloading 30,000 individual author webpages. However, the latter already contain the data (much more of it, 367,478 papers) required in the regression in Table 6. We give the results for this regression based on the author webpage data in Table 21, which are qualitatively similar to the results in Table 6, and we still have the empirical formula (4).[24] As mentioned in Table 17, the Top Papers data sample is actually small, despite the 9,999 papers it contains, so any results obtained using the Top Papers data should be taken with a grain of salt.

Let us now discuss possible variations and generalizations of our indexes – and this topic invariably overlaps with caveats. E.g., as in the case of the $h$-index, typical values of our indexes – naturally – will vary from discipline to discipline. One way to deal with this is to simply compute the indexes separately for each discipline. As mentioned above, SSRN provides the Top Authors (but not the Top Papers) data broken down by some disciplines (law, business and economics), with such data for other disciplines (to wit, accounting and finance) apparently forthcoming. It would be interesting to statistically analyze our indexes by each discipline.[25]

One way the disparity in the $h$-index across different disciplines has been dealt with is via normalizing the $h$-index (or the citations) within each discipline (e.g., via dividing by a mean for the discipline).[26] A similar approach can be applied to our indexes as well. So, can we simply rescale SSRN downloads by some factor and apply the $h$-index to the rescaled downloads (as, e.g., for the internet media in (Hovden, 2013))? A natural choice for such a factor is, e.g., the median of $z = d_{tot} / n_{tot}^2$ (the distribution is too skewed to use the mean, which is much higher), which is median$(z) \approx 29.07$ (based on the Top Authors data). However, if we apply the $h$-index to $d' = d_{tot} / 30$, expectedly, we get an index with a highly skewed distribution.

---

[24] With $r$ replaced by $r_p$ and $d_{tot}$ replaced by $d_p$ (see Table 21). The inflection point in the curve $r_p$ v. $\ln(d_p)$ occurs at around $d_p \approx 70$. Let us note that the author webpage data contains the numbers of overall downloads for each paper by each of the 30,000 authors, but not for the last 12 months. So, we have no choice but to use to the Top Papers data for the regressions in Tables 7 and 8.

[25] We have refrained from doing so here as the breakdown by all disciplines is currently unavailable. Let us note that the author webpage data can also be split by discipline as the SSRN IDs are obtained via the Top Authors data.

[26] See, e.g., (Anauati, Galiani, and Gálvez, 2014), (Batista et al, 2006), (Bornmann and Daniel, 2008), (Iglesias and Pecharromán, 2007), (Kaur, Radicchi and Menczer, 2013), (Podlubny, 2005), (Podlubny and Kassayova, 2006).

There is no escaping the fact that SSRN downloads are exponential by nature as we – hopefully convincingly – argued above based on the empirical data. Logs of the numbers of downloads – not the numbers of downloads – are a natural measure. And this is essentially our key observation. Once we accept this fact, the indexes we propose are natural, with possible tweaks (e.g., the linear extrapolation between the points $(k, \ln(d_k))$ and $(k+1, \ln(d_{k+1}))$ in Eq. (3) can be tweaked, but this is all minutiae). Just as the $h$-index, our indexes use only each author's data, but no cross-sectional (across a sample of authors) data, which makes them easy to compute. Rescaling $d_{tot}$ by some factor that requires cross-sectional data to compute would complicate the calculation of an index. And, once again, such rescaling does nothing to address the exponential nature of SSRN downloads, while our indexes are built on that very premise.

Several variations of and indexes complementary to the $h$-index have been proposed, e.g., the $g$-index (Egghe, 2006), whose analog for SSRN downloads is our $\kappa_*$ index. A geometric mean of the $h$-index and the $g$-index – the so-called $hg$-index (Alonso et al., 2010) – is a composite index (also, see, e.g., (Franceschini and Maisano, 2011)). An evident application to our indexes would be to consider geometric means of the $k$ and $\kappa$ or $k_*$ and $\kappa_*$ indexes.

Another point worth mentioning relates to the overall normalization factor $\gamma$ we introduced in Subsection 3.4. In Eq. (12) it is implicitly assumed to be a constant. However, there is no reason (other than forgoing simplicity, that is) why we could not consider non-constant $\gamma(d_i)$ in Eq. (12) instead, where $\gamma(\cdot)$ is some – in many cases, likely relatively slowly varying – function.

Self-citations affect the total number of citations as well as the $h$-index. For citations this is relatively simple to deal with: one can simply remove self-citations. E.g., INSPIRE provides such functionality. Analogously, self-downloads can be a nuisance for SSRN downloads (see, e.g., (Edelman and Larkin, 2014)) and thereby our indexes. Dealing with self-downloads is harder, perhaps even internally at SSRN. This is a caveat. Also, some papers are not available for download from SSRN (due to journal policies – see above – with exceptions often granted to established authors). Naturally, as with any index, there are caveats. Nonetheless, our indexes are novel and it would be interesting if SSRN could analyze and perhaps even implement them.

## Acknowledgments

I would like to thank Ludo Waltman for reading a draft of the manuscript and valuable comments and suggestions that have helped improve it, and Blaise Cronin for encouragement.

# Tables

| Author Name, SSRN ID | $k_*$ | $k$ | $d_*$ | Total # of Downloads | # $n$ of $d_i > 0$ Papers | SSRN Rank |
|---|---|---|---|---|---|---|
| Michael C. Jensen, 9 | 9.959 | 9 | 21154 | 828131 | 122 | 1 |
| Eugene F. Fama, 998 | 9.538 | 9 | 13884 | 421883 | 37 | 3 |
| Pablo Fernandez, 12696 | 9.476 | 9 | 13047 | 813639 | 230 | 2 |
| Kenneth R. French, 1455 | 9.030 | 9 | 8350 | 265339 | 34 | 4 |
| Attilio Meucci, 403805 | 8.833 | 8 | 6862 | 172719 | 38 | 15 |
| Aswath Damodaran, 20838 | 8.773 | 8 | 6460 | 177761 | 48 | 14 |
| William N. Goetzmann, 2309 | 8.747 | 8 | 6292 | 221081 | 97 | 9 |
| Lucian A. Bebchuk, 17037 | 8.700 | 8 | 6008 | 232806 | 121 | 8 |
| Shahin Shojai, 342721 | 8.608 | 8 | 5478 | 90605 | 19 | 47 |
| Kostas Koufopoulos, 358994 | 8.598 | 8 | 5425 | 126117 | 19 | 30 |
| Andrew W. Lo, 17399 | 8.593 | 8 | 5394 | 146789 | 75 | 21 |
| Daniel Kaufmann, 163813 | 8.581 | 8 | 5333 | 193401 | 51 | 12 |
| Nassim Nicholas Taleb, 475810 | 8.571 | 8 | 5278 | 147993 | 16 | 20 |
| Werner Erhard, 433651 | 8.563 | 8 | 5236 | 141282 | 24 | 22 |
| Christian Leuz, 18004 | 8.500 | 8 | 4914 | 131136 | 35 | 26 |
| Stephen H. Penman, 15046 | 8.466 | 8 | 4750 | 127624 | 33 | 29 |
| Bernard S. Black, 16042 | 8.416 | 8 | 4521 | 163455 | 129 | 19 |
| Aart Kraay, 42707 | 8.412 | 8 | 4504 | 167904 | 56 | 17 |
| Ignacio Velez-Pareja, 145648 | 8.393 | 8 | 4418 | 254252 | 215 | 5 |
| Campbell R. Harvey, 16198 | 8.384 | 8 | 4377 | 201786 | 116 | 11 |

**Table 1.** Top 20 SSRN authors by the index $k_*$. The 6th column is the number of papers with at least 1 download. The 2nd column is rounded down to the 3rd decimal. The 4th column is rounded down to the nearest integer. The SSRN rank is based on the total number of downloads. All statistics are as of the date(s) of our downloads of the data (see Section 2).

| Quantity | Min. | 1st Quartile | Median | Mean | 3rd Quartile | Max. |
|---|---|---|---|---|---|---|
| $d_{tot}$ (overall) | 129 | 672 | 1626 | 3844 | 3757 | 824800 |
| $\ln(d_{tot})$ (overall) | 4.86 | 6.51 | 7.394 | 7.406 | 8.231 | 13.62 |
| $n_{tot}$ (overall) | 1 | 3 | 8 | 13.4 | 16 | 365 |
| $\ln(n_{tot})$ (overall) | 0 | 1.099 | 2.079 | 2.001 | 2.773 | 5.9 |
| $d_{tot}/n_{tot}$ (overall) | 2 | 113 | 210 | 403.1 | 418 | 37340 |
| $\ln(d_{tot}/n_{tot})$ (overall) | 0.6931 | 4.727 | 5.347 | 5.406 | 6.035 | 10.53 |
| $d_{tot}$ (12 mo) | 129 | 177 | 271 | 527.1 | 498 | 139300 |
| $\ln(d_{tot})$ (12 mo) | 4.86 | 5.176 | 5.602 | 5.798 | 6.211 | 11.84 |
| $n_{tot}$ (12 mo) | 1 | 1 | 2 | 2.612 | 3 | 193 |
| $\ln(n_{tot})$ (12 mo) | 0 | 0 | 0.6931 | 0.6441 | 1.099 | 5.263 |
| $d_{tot}/n_{tot}$ (12 mo) | 2 | 113 | 210 | 403.1 | 418 | 37340 |
| $\ln(d_{tot}/n_{tot})$ (12 mo) | 0.6931 | 4.727 | 5.347 | 5.406 | 6.035 | 10.53 |
| $d_{tot}^{overall}/d_{tot}^{12-mo}$ | 1 | 2.55 | 5.172 | 7.354 | 9.973 | 169 |
| $\ln(d_{tot}^{overall}/d_{tot}^{12-mo})$ | 0 | 0.9362 | 1.643 | 1.608 | 2.3 | 5.13 |

**Table 2.** Cross-sectional (across all authors in the SSRN Top Authors data) summaries for the total number of downloads $d_{tot}$ and its log, the total number of papers $n_{tot}$ and its log, and downloads-per-paper $d_{tot}/n_{tot}$ and its log, both overall and for the last 12 months. In $n_{tot}$ for the last 12 months we drop all $n_{tot} = 0$ cases. We give the numbers as rounded by R. E.g., the maximum numbers of downloads overall and in the last 12 months actually are 824762 and 139290, respectively. The $d_{tot}/n_{tot}$ figures are already rounded to the nearest integer in the SSRN Top Authors data. The bottom two rows summarize the ratio of the overall total number of downloads to the 12-mo total number of downloads and the log of this ratio.

| Regression: $\ln(r) \sim \ln(d_{tot}) + \ln^2(d_{tot})$ | Estimate | Standard error | t-statistic | Overall statistics |
|---|---|---|---|---|
| Intercept | 4.704 | 0.00626 | 751.8 | |
| $\ln(d_{tot})$ | 2.009 | 0.00167 | 1203.7 | |
| $\ln^2(d_{tot})$ | -0.182 | 0.00011 | -1661.9 | |
| Multiple/Adjusted R-squared | | | | 0.9985 |
| F-statistic | | | | $9.771 \times 10^6$ |

**Table 3.** Summary (using the function `summary(lm())` in R) for the cross-sectional (over all authors in the SSRN Top Authors data) polynomial regression of $\ln(r)$ over $\ln(d_{tot})$ and $\ln^2(d_{tot})$ with the intercept. The regression formula reads `lm(y ~ x + I(x^2))` in R notations, where $y = \ln(r)$ and $x = \ln(d_{tot})$. Here the rank $r$ and $d_{tot}$ are based on the overall downloads. We keep the outliers in the regression not to inflate the statistics.

| Regression: $\ln(r) \sim \ln(d_{tot}) + \ln^2(d_{tot})$ | Estimate | Standard error | t-statistic | Overall statistics |
|---|---|---|---|---|
| Intercept | 11.18 | 0.00590 | 1893.1 | |
| $\ln(d_{tot})$ | 0.492 | 0.00189 | 260.3 | |
| $\ln^2(d_{tot})$ | -0.138 | 0.00015 | -924.7 | |
| Multiple/Adjusted R-squared | | | | 0.9994 |
| F-statistic | | | | $2.413 \times 10^7$ |

**Table 4.** Same as Table 3 with the rank $r$ and $d_{tot}$ based on the last 12-months downloads.

| Quantity | Min. | 1st Quartile | Median | Mean | 3rd Quartile | Max. |
|---|---|---|---|---|---|---|
| $d_p$ (overall) | 160 | 370 | 805 | 1800 | 1834 | 152200 |
| $\ln(d_p)$ (overall) | 5.075 | 5.914 | 6.691 | 6.785 | 7.514 | 11.93 |
| $d_p$ (12 mo) | 160 | 191.8 | 247 | 384.4 | 373 | 31860 |
| $\ln(d_p)$ (12 mo) | 5.075 | 5.256 | 5.509 | 5.68 | 5.922 | 10.37 |

**Table 5.** Cross-sectional (across all papers in the SSRN Top Papers data) summaries for the number of downloads $d_p$ and its log, both overall and for the last 12 months. We give the numbers as rounded by R. E.g., the maximum numbers of downloads overall and in the last 12 months actually are 152242 and 31862, respectively.

| Regression: $\ln(r_p) \sim \ln(d_p) + \ln^2(d_p)$ | Estimate | Standard error | t-statistic | Overall statistics |
|---|---|---|---|---|
| Intercept | 4.375 | 0.01208 | 362.1 | |
| $\ln(d_p)$ | 1.952 | 0.00345 | 566.0 | |
| $\ln^2(d_p)$ | -0.199 | 0.00024 | -822.9 | |
| Multiple/Adjusted R-squared | | | | 0.9986 |
| F-statistic | | | | $3.602 \times 10^6$ |

**Table 6.** Summary (using the function `summary(lm())` in R) for the cross-sectional (over all papers in the SSRN Top Papers data) polynomial regression of $\ln(r_p)$ over $\ln(d_p)$ and $\ln^2(d_p)$ with the intercept. The regression formula reads `lm(y ~ x + I(x^2))` in R notations, where $y = \ln(r_p)$ and $x = \ln(d_p)$. Here the rank $r_p$ and $d_p$ are based on the overall downloads. We keep the outliers in the regression not to inflate the statistics.

| Regression: $\ln(r_p) \sim \ln(d_p) + \ln^2(d_p)$ | Estimate | Standard error | t-statistic | Overall statistics |
|---|---|---|---|---|
| Intercept | 16.78 | 0.01772 | 947.0 | |
| $\ln(d_p)$ | -1.331 | 0.00574 | -232.1 | |
| $\ln^2(d_p)$ | -0.031 | 0.00046 | -67.06 | |
| Multiple/Adjusted R-squared | | | | 0.9991 |
| F-statistic | | | | $5.494 \times 10^6$ |

**Table 7.** Same as Table 6 with the rank $r_p$ and $d_p$ based on the last 12-months downloads.

| Regression: $\ln(r_p) \sim \ln(d_p)$ | Estimate | Standard error | t-statistic | Overall statistics |
|---|---|---|---|---|
| Intercept | 17.95 | 0.00356 | 5047 | |
| $\ln(d_p)$ | -1.715 | 0.00062 | -2753 | |
| Multiple/Adjusted R-squared | | | | 0.9987 |
| F-statistic | | | | $7.576 \times 10^6$ |

**Table 8.** Summary (using the function `summary(lm())` in R) for the cross-sectional (over all papers in the SSRN Top Papers data) linear regression of $\ln(r_p)$ over $\ln(d_p)$ with the intercept. The regression formula reads `lm(y ~ x)` in R notations, where $y = \ln(r_p)$ and $x = \ln(d_p)$. Here the rank $r_p$ and $d_p$ are based on the last 12-months downloads. We keep the outliers in the regression not to inflate the statistics.

| Quantity | Min. | 1st Quartile | Median | Mean | 3rd Quartile | Max. |
|---|---|---|---|---|---|---|
| $k_*$ | 0 | 3.432 | 4.484 | 4.32 | 5.334 | 9.96 |
| $k / n_{tot}$ | 0.01049 | 0.3077 | 0.5556 | 0.5841 | 1 | 1 |
| $w$ | 1 | 1.878 | 2.631 | 3.407 | 3.6 | 213.5 |
| $\ln(w)$ | 0 | 0.63 | 0.9675 | 0.9718 | 1.281 | 5.364 |
| $v = \ln(d_{tot} / n_{tot}) / n_{tot}$ | 0.00539 | 0.336 | 0.6896 | 1.502 | 1.673 | 10.41 |

**Table 9.** Cross-sectional (across all authors in the SSRN Top Authors data) summaries for $k_*$ and the ratio $k / n_{tot}$ together with the factor $w$ (see Eq. (7)), its log and $v$. NAs are omitted.

| Quantity | Min. | 1st Quartile | Median | Mean | 3rd Quartile | Max. |
|---|---|---|---|---|---|---|
| $\kappa_*$ | 0 | 3.752 | 5.197 | 4.862 | 6.024 | 10.47 |
| $\kappa / n_{tot}$ | 0.01049 | 0.3333 | 0.6667 | 0.6474 | 1 | 1 |
| $\omega$ | 1 | 2.023 | 2.858 | 8.757 | 4.109 | 12460 |
| $\ln(\omega)$ | 0 | 0.7048 | 1.05 | 1.151 | 1.413 | 9.431 |

**Table 10.** Same as Table 9 for the indexes $\kappa$ and $\kappa_*$ (except there is no nontrivial analog of $v$ in this case – see Subsection 3.1). The factor $\omega$ is defined in Subsection 3.1.

| Author Name, SSRN ID | $\kappa_*$ | $\kappa$ | $f_*$ | Total # of Downloads | # $n$ of $d_i > 0$ Papers | SSRN Rank |
|---|---|---|---|---|---|---|
| Michael C. Jensen, 9 | 10.468 | 10 | 35185 | 828131 | 122 | 1 |
| Eugene F. Fama, 998 | 10.273 | 10 | 28955 | 421883 | 37 | 3 |
| Pablo Fernandez, 12696 | 9.979 | 9 | 21576 | 813639 | 230 | 2 |
| Daniel J. Solove, 249137 | 9.870 | 9 | 19359 | 233202 | 42 | 7 |
| Kenneth R. French, 1455 | 9.841 | 9 | 18795 | 265339 | 34 | 4 |
| William H. Meckling, 896 | 9.785 | 9 | 17776 | 178280 | 17 | 13 |
| Daniel Kaufmann, 163813 | 9.657 | 9 | 15638 | 193401 | 51 | 12 |
| Aart Kraay, 42707 | 9.636 | 9 | 15318 | 167904 | 56 | 17 |
| Massimo Mastruzzi, 332182 | 9.575 | 9 | 14405 | 137898 | 10 | 24 |
| Nassim Nicholas Taleb, 475810 | 9.551 | 9 | 14067 | 147993 | 16 | 20 |
| Andrew Metrick, 20387 | 9.442 | 9 | 12614 | 134101 | 68 | 25 |
| Werner Erhard, 433651 | 9.423 | 9 | 12379 | 141282 | 24 | 22 |
| John R. Lott, 16317 | 9.380 | 9 | 11855 | 129086 | 32 | 27 |
| Mathew O. Jackson, 161894 | 9.369 | 9 | 11730 | 121642 | 45 | 32 |
| Kostas Koufopoulos, 358994 | 9.359 | 9 | 11609 | 126117 | 19 | 30 |
| Stephen H. Penman, 15046 | 9.350 | 9 | 11503 | 127624 | 33 | 29 |
| William N. Goetzmann, 2309 | 9.302 | 9 | 10964 | 221081 | 97 | 9 |
| Attilio Meucci, 403805 | 9.285 | 9 | 10781 | 172719 | 38 | 15 |
| Aswath Damodaran, 20838 | 9.259 | 9 | 10508 | 177761 | 48 | 14 |
| K. Geert Rouwenhorst, 59592 | 9.247 | 9 | 10373 | 109747 | 20 | 36 |

**Table 11.** Top 20 SSRN authors by the index $\kappa_*$. The 6th column is the number of papers with at least 1 download. The 2nd column is rounded down to the 3rd decimal. The 4th column is rounded down to the nearest integer. The SSRN rank is based on the total number of downloads. All statistics are as of the date(s) of our downloads of the data (see Section 2).

| Regression: $k_* \sim \ln(d_{tot})$ | Estimate | Standard error | t-statistic | Overall statistics |
|---|---|---|---|---|
| Intercept | -2.920 | 0.03046 | -95.87 | |
| $\ln(d_{tot})$ | 0.964 | 0.00405 | 237.99 | |
| Multiple/Adjusted R-squared | | | | 0.6539 |
| F-statistic | | | | $5.664 \times 10^4$ |

**Table 12.** Summary (using the function `summary(lm())` in R) for the cross-sectional (over all authors in the SSRN Top Authors data) linear regression of $k_*$ over $\ln(d_{tot})$ with the intercept.

| Regression: $k \sim \ln(d_{tot})$ | Estimate | Standard error | t-statistic | Overall statistics |
|---|---|---|---|---|
| Intercept | -3.803 | 0.03520 | -108.0 | |
| $\ln(d_{tot})$ | 1.006 | 0.00468 | 214.9 | |
| Multiple/Adjusted R-squared | | | | 0.6064 |
| F-statistic | | | | $4.618 \times 10^4$ |

**Table 13.** Summary (using the function `summary(lm())` in R) for the cross-sectional (over all authors in the SSRN Top Authors data) linear regression of $k$ over $\ln(d_{tot})$ with the intercept.

| Regression: $\kappa_* \sim \ln(d_{tot})$ | Estimate | Standard error | t-statistic | Overall statistics |
|---|---|---|---|---|
| Intercept | -3.319 | 0.03723 | -89.13 | |
| $\ln(d_{tot})$ | 1.090 | 0.00495 | 220.12 | |
| Multiple/Adjusted R-squared | | | | 0.6178 |
| F-statistic | | | | $4.845 \times 10^4$ |

**Table 14.** Summary (using the function `summary(lm())` in R) for the cross-sectional (over all authors in the SSRN Top Authors data) linear regression of $\kappa_*$ over $\ln(d_{tot})$ with the intercept.

| Regression: $\kappa \sim \ln(d_{tot})$ | Estimate | Standard error | t-statistic | Overall statistics |
|---|---|---|---|---|
| Intercept | -4.327 | 0.04215 | -102.7 | |
| $\ln(d_{tot})$ | 1.145 | 0.00560 | 204.3 | |
| Multiple/Adjusted R-squared | | | | 0.582 |
| F-statistic | | | | $4.174 \times 10^4$ |

**Table 15.** Summary (using the function `summary(lm())` in R) for the cross-sectional (over all authors in the SSRN Top Authors data) linear regression of $\kappa$ over $\ln(d_{tot})$ with the intercept.

| Regression: $X \sim \ln(d_{tot}) + \ln(n_{tot})$ | $X = k_*$ | $X = k$ | $X = \kappa_*$ | $X = \kappa$ |
|---|---|---|---|---|
| Estimate: Intercept | -0.887 | -1.498 | -0.914 | -1.630 |
| Estimate: $\ln(d_{tot})$ | 0.499 | 0.479 | 0.540 | 0.528 |
| Estimate: $\ln(n_{tot})$ | 0.749 | 0.849 | 0.886 | 0.994 |
| Standard Error: Intercept | 0.02234 | 0.02646 | 0.02845 | 0.03255 |
| Standard Error: $\ln(d_{tot})$ | 0.00352 | 0.00417 | 0.00448 | 0.00513 |
| Standard Error: $\ln(n_{tot})$ | 0.00374 | 0.00443 | 0.00476 | 0.00544 |
| t-statistic: Intercept | -39.7 | -56.62 | -32.12 | -50.08 |
| t-statistic: $\ln(d_{tot})$ | 141.8 | 114.88 | 120.48 | 103.04 |
| t-statistic: $\ln(n_{tot})$ | 200.5 | 191.89 | 186.25 | 182.57 |
| Multiple/Adjusted R-squared | 0.8522 | 0.8234 | 0.8228 | 0.8021 |
| F-statistic | $8.642 \times 10^4$ | $6.986 \times 10^4$ | $6.960 \times 10^4$ | $6.074 \times 10^4$ |

**Table 16.** Summaries (using the function `summary(lm())` in R) for the cross-sectional (over all authors in the SSRN Top Authors data) linear regressions of $X$ over the two explanatory variables $\ln(d_{tot})$ and $\ln(n_{tot})$ with the intercept, where $X = k_*$, $k$, $\kappa_*$, $\kappa$. Cf. Tables 12-15.

| Quantity | Min. | 1st Quartile | Median | Mean | 3rd Quartile | Max. |
|---|---|---|---|---|---|---|
| $k_*$ (12 months) (all) | 1.803 | 1.811 | 1.823 | 2.246 | 1.865 | 7.966 |
| $\kappa_*$ (12 months) (all) | 1.803 | 1.811 | 1.823 | 2.272 | 1.865 | 8.505 |
| $k_*$ (12 months) ($n_{tot} > 1$) | 2.755 | 2.767 | 2.797 | 3.605 | 4.530 | 7.966 |
| $\kappa_*$ (12 months) ($n_{tot} > 1$) | 2.755 | 2.816 | 2.780 | 3.710 | 4.599 | 8.505 |
| $k_*$ (12 months) ($n_{tot} > 2$) | 3.675 | 3.696 | 4.545 | 4.574 | 5.272 | 7.966 |
| $\kappa_*$ (12 months) ($n_{tot} > 2$) | 3.682 | 3.738 | 4.623 | 4.786 | 5.608 | 8.505 |

**Table 17.** Cross-sectional (across all authors in the SSRN Top Papers data) summaries for the $k_*$ and $\kappa_*$ indexes based on the last 12-months downloads obtained using the Top Papers data. Since we only have 9,999 papers with 11,871 authors (see Subsection 3.3), i.e., the data sample is in fact small despite a large number of papers, the data mostly has authors with only 1 or 2 papers. This causes the bulk of the index values to be artificially low (and the primary cause of this is not the fact that the bulk of the 12-mo download numbers is lower than the bulk of the overall download numbers by roughly a factor of 5-7; see the bottom two rows in Table 2). Therefore, we provide summaries for all 11,871 authors, the 2,834 authors with $n_{tot} > 1$ papers, and the 1,310 authors with $n_{tot} > 2$ papers.

| Author Name, SSRN ID | $k_*$ | $k$ | $d_*$ | Total # of Downloads (12 mo) | # $n$ of Papers (12 mo) | Downloads Rank (12 mo) |
|---|---|---|---|---|---|---|
| Pablo Fernandez, 12696 | 7.965 | 7 | 2881 | 136064 | 114 | 1 |
| Michael C. Jensen, 9 | 7.714 | 7 | 2240 | 62397 | 62 | 2 |
| Aswath Damodaran, 20838 | 7.193 | 7 | 1331 | 34153 | 30 | 5 |
| Attilio Meucci, 403805 | 7.162 | 7 | 1289 | 25724 | 32 | 14 |
| Eugene F. Fama, 998 | 7.074 | 7 | 1180 | 31295 | 28 | 8 |
| Tobias J. Moskowitz, 189945 | 7.056 | 7 | 1160 | 27283 | 13 | 11 |
| Campbell R. Harvey, 16198 | 7.036 | 7 | 1137 | 31193 | 38 | 10 |
| Werner Erhard, 433651 | 7.001 | 7 | 1097 | 22622 | 26 | 16 |
| Isabel Fernández Acín, 2039798 | 6.744 | 6 | 849 | 18895 | 11 | 19 |
| Kenneth R. French, 1455 | 6.705 | 6 | 816 | 25890 | 23 | 12 |
| Lasse Heje Pedersen, 277060 | 6.663 | 6 | 783 | 15659 | 10 | 24 |
| Dan M. Kahan, 45442 | 6.615 | 6 | 746 | 14905 | 23 | 26 |
| Matthew O. Jackson, 161894 | 6.605 | 6 | 739 | 39857 | 15 | 3 |
| Cass R. Sunstein, 16333 | 6.566 | 6 | 710 | 25875 | 46 | 13 |
| George Serafeim, 573672 | 6.541 | 6 | 693 | 11526 | 17 | 36 |
| Clifford S. Asness, 77768 | 6.504 | 6 | 668 | 23771 | 12 | 15 |
| John R. Graham, 17209 | 6.493 | 6 | 660 | 12275 | 28 | 32 |
| Kari L. Granger, 805286 | 6.471 | 6 | 646 | 11913 | 9 | 35 |
| Guofu Zhou, 16976 | 6.470 | 6 | 645 | 8428 | 14 | 56 |
| Wade D. Pfau, 388906 | 6.462 | 6 | 640 | 15877 | 13 | 22 |

**Table 18.** Top 20 SSRN authors by the index $k_*$ based on the last 12-months downloads obtained using the Top Papers data. See Table 1 for number rounding and other information.

| Author Name, SSRN ID | $\kappa_*$ | $\kappa$ | $f_*$ | Total # of Downloads (12 mo) | # $n$ of Papers (12 mo) | Downloads Rank (12 mo) |
|---|---|---|---|---|---|---|
| Pablo Fernandez, 12696 | 8.504 | 8 | 4938 | 136064 | 114 | 1 |
| Matthew O. Jackson, 161894 | 8.414 | 8 | 4509 | 39857 | 15 | 3 |
| Mebane T. Faber, 649342 | 8.309 | 8 | 4060 | 37296 | 8 | 4 |
| Michael C. Jensen, 9 | 8.173 | 8 | 3547 | 62397 | 62 | 2 |
| Aswath Damodaran, 20838 | 8.082 | 8 | 3237 | 34153 | 30 | 5 |
| Tobias J. Moskowitz, 189945 | 8.052 | 8 | 3142 | 27283 | 13 | 11 |
| Eugene F. Fama, 998 | 7.984 | 7 | 2935 | 31295 | 28 | 8 |
| Clifford S. Asness, 77768 | 7.954 | 7 | 2848 | 23771 | 12 | 15 |
| Kenneth R. French, 1455 | 7.903 | 7 | 2706 | 25890 | 23 | 12 |
| Campbell R. Harvey, 16198 | 7.831 | 7 | 2518 | 31193 | 38 | 10 |
| Isabel Fernández Acín, 2039798 | 7.706 | 7 | 2222 | 18895 | 11 | 19 |
| Pablo Linares, 1907849 | 7.605 | 7 | 2009 | 15557 | 9 | 25 |
| Lasse Heje Pedersen, 277060 | 7.566 | 7 | 1933 | 15659 | 10 | 24 |
| Wade D. Pfau, 388906 | 7.563 | 7 | 1926 | 15877 | 13 | 22 |
| Werner Erhard, 433651 | 7.541 | 7 | 1883 | 22622 | 26 | 16 |
| Andrea Frazzini, 384604 | 7.495 | 7 | 1799 | 20489 | 7 | 17 |
| Daniel J. Solove, 249137 | 7.491 | 7 | 1793 | 15927 | 17 | 21 |
| Cass R. Sunstein, 16333 | 7.426 | 7 | 1679 | 25875 | 46 | 13 |
| Attilio Meucci, 403805 | 7.398 | 7 | 1632 | 25724 | 32 | 14 |
| Kari L. Granger, 805286 | 7.358 | 7 | 1569 | 11913 | 9 | 35 |

**Table 19.** Top 20 SSRN authors by the index $\kappa_*$ based on the last 12-months downloads obtained using the Top Papers data. See Table 1 for number rounding and other information.

| Regression: $X \sim \ln(T)$ | $X = \ln(d_{tot})$ | $X = \ln(d_{tot} / n_{tot})$ | $X = k_*$ | $X = \kappa_*$ |
|---|---|---|---|---|
| Estimate: Intercept | 5.370 | 4.893 | 2.172 | 2.435 |
| Estimate: $\ln(T)$ | 1.126 | 0.316 | 1.182 | 1.339 |
| Standard Error: Intercept | 0.01380 | 0.01602 | 0.01689 | 0.01956 |
| Standard Error: $\ln(T)$ | 0.00677 | 0.00786 | 0.00829 | 0.00960 |
| t-statistic: Intercept | 389.2 | 305.42 | 128.6 | 124.5 |
| t-statistic: $\ln(T)$ | 166.2 | 40.22 | 142.6 | 139.5 |
| Multiple/Adjusted R-squared | 0.4918 | 0.0536 | 0.416 | 0.4052 |
| F-statistic | $2.763 \times 10^4$ | 1618 | $2.034 \times 10^4$ | $1.945 \times 10^4$ |

**Table 20.** Summaries (using the function `summary(lm())` in R) for the cross-sectional (over all authors in the SSRN Top Authors data with $T \geq 1$) linear regressions of $X$ over $\ln(T)$ with the intercept, where $X = \ln(d_{tot})$, $\ln(d_{tot} / n_{tot})$, $k_*$, $\kappa_*$. Here $T$ is the time in (years) from the date of the author's first posting of a paper ("Posted:" field) on SSRN until August 17, 2015.

| Regression: $\ln(r_p) \sim \ln(d_p) + \ln^2(d_p)$ | Estimate | Standard error | t-statistic | Overall statistics |
|---|---|---|---|---|
| Intercept | 12.17 | $1.069 \times 10^{-3}$ | 11381 | |
| $\ln(d_p)$ | 0.612 | $4.711 \times 10^{-4}$ | 1298 | |
| $\ln^2(d_p)$ | -0.133 | $5.105 \times 10^{-5}$ | -2607 | |
| Multiple/Adjusted R-squared | | | | 0.9882 |
| F-statistic | | | | $1.539 \times 10^7$ |

**Table 21.** Same as Table 6, except all quantities are based on the author webpage data.

# Figures

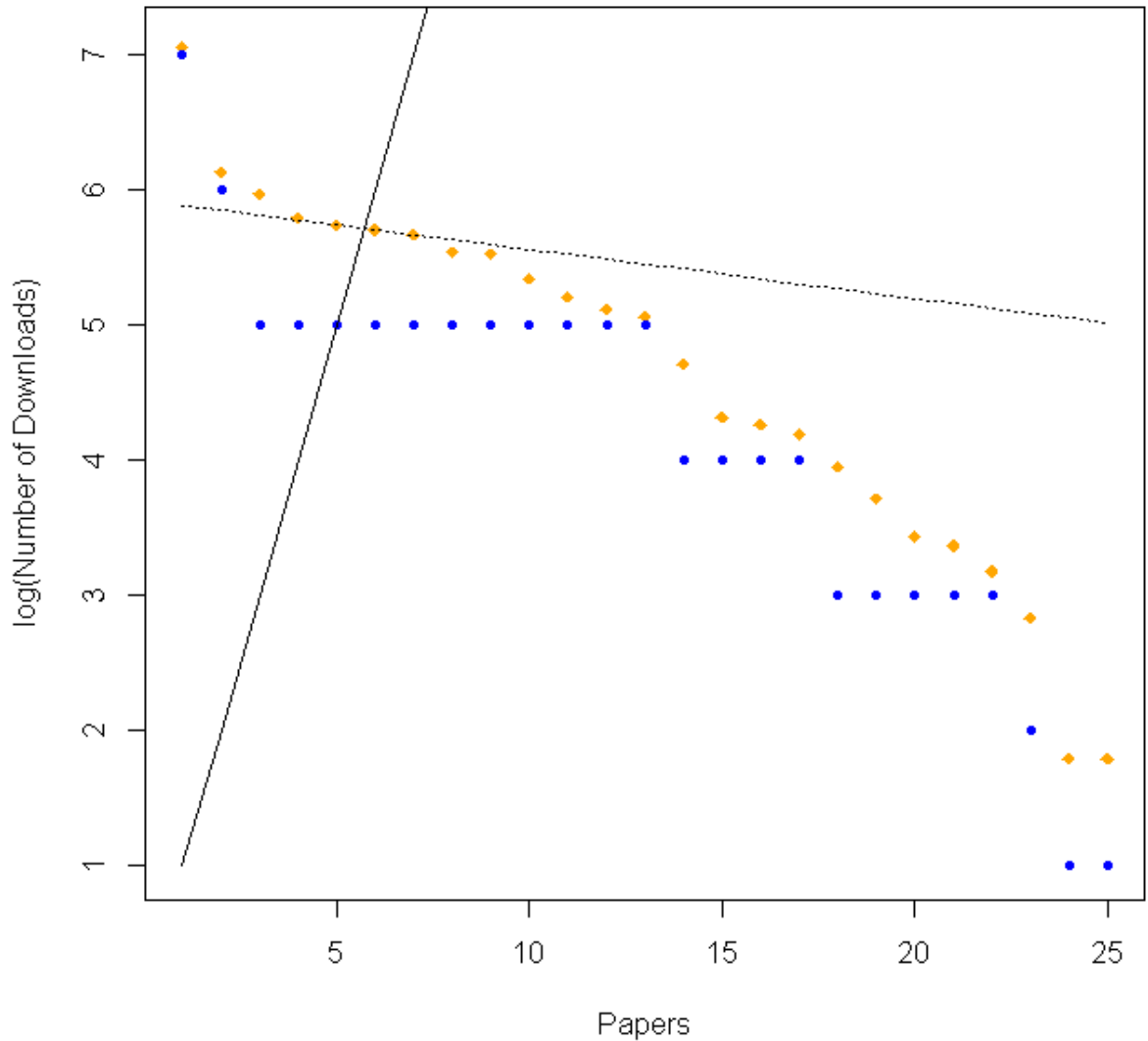

**Figure 1.** This figure illustrates the computation of the $k$, $k_*$ and $d_*$ indexes for a randomly chosen author. The sloping diamonds correspond to $\ln(d_i)$. The horizontal circles correspond to $q_i = \lfloor \ln(d_i) \rfloor$ in Eq. (2). The solid straight line has slope 1 and its intersection with the horizontal lines of circles gives the value of $k = 5$. The dotted straight line with the negative slope goes through the points $(k, \ln(d_k))$ and $(k+1, \ln(d_{k+1}))$ and its intersection with the solid straight line determines $k_*$. In this example we have $k_* = 5.710$ and $d_* = 302$.

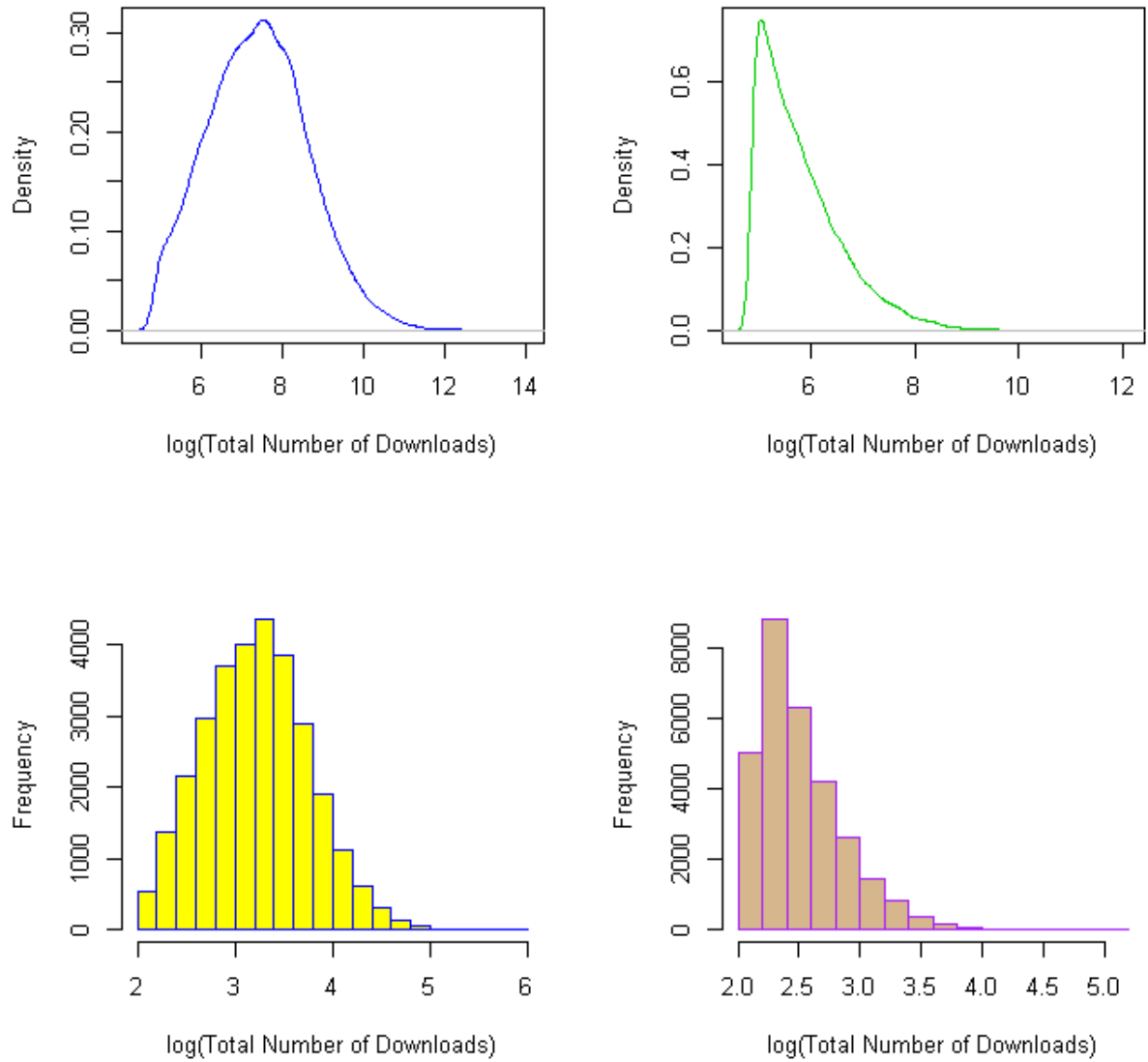

**Figure 2.** Cross-sectional (across all authors in the SSRN Top Authors data) density for $\ln(d_{tot})$ and histogram for $\log_{10}(d_{tot})$, where $d_{tot}$ is the total number of downloads for each author. Left column: overall; right column: the last 12 months.

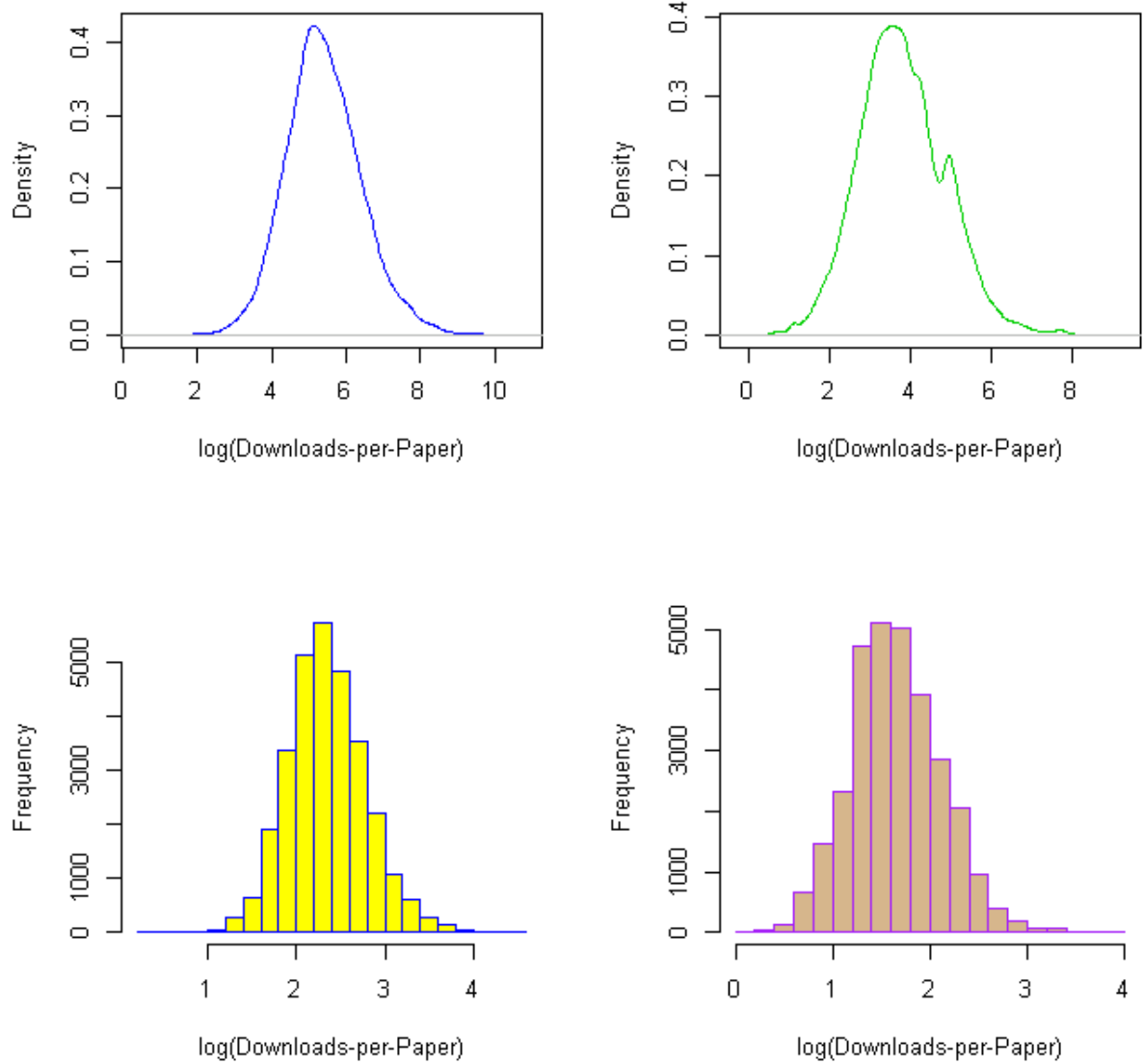

**Figure 3.** Cross-sectional (across all authors in the SSRN Top Authors data) density for $\ln(d_{tot} / n_{tot})$ and histogram for $\log_{10}(d_{tot} / n_{tot})$, where $d_{tot}$ is the total number of downloads for each author, while $n_{tot}$ is the number of papers. Left column: overall; right column: the last 12 months.

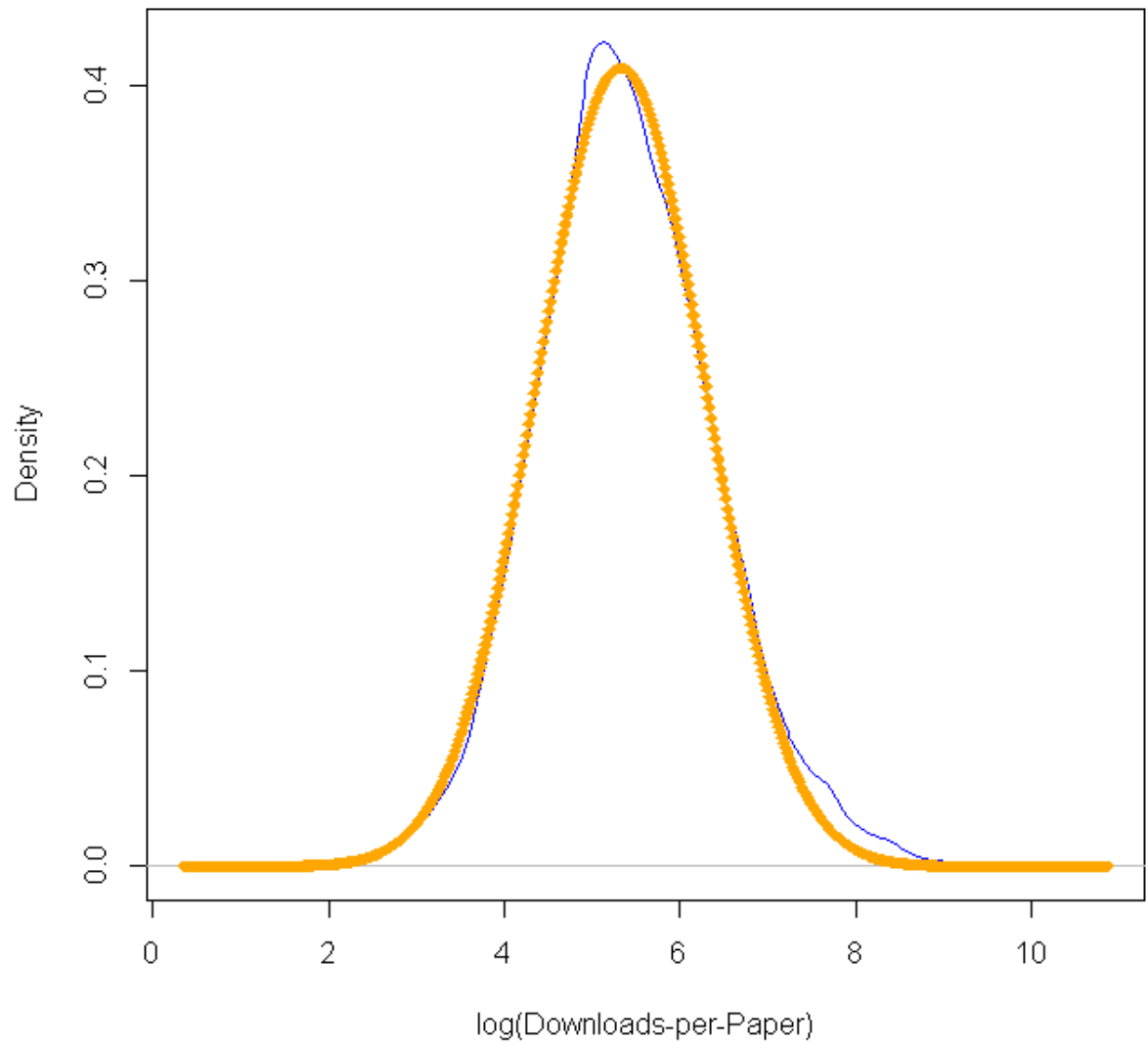

**Figure 4.** The solid line (not a Gaussian) is the same as the upper-left density curve in Figure 3 (mean = 5.406 and standard deviation = 1.016 based on $\ln(d_{tot} / n_{tot})$, with maximum value = 0.423 based on the density) for overall downloads. The diamonds correspond to the least-squares fit Gaussian curve (mean = 5.329, standard deviation = 0.961, maximum value = 0.409, all based on the fit).

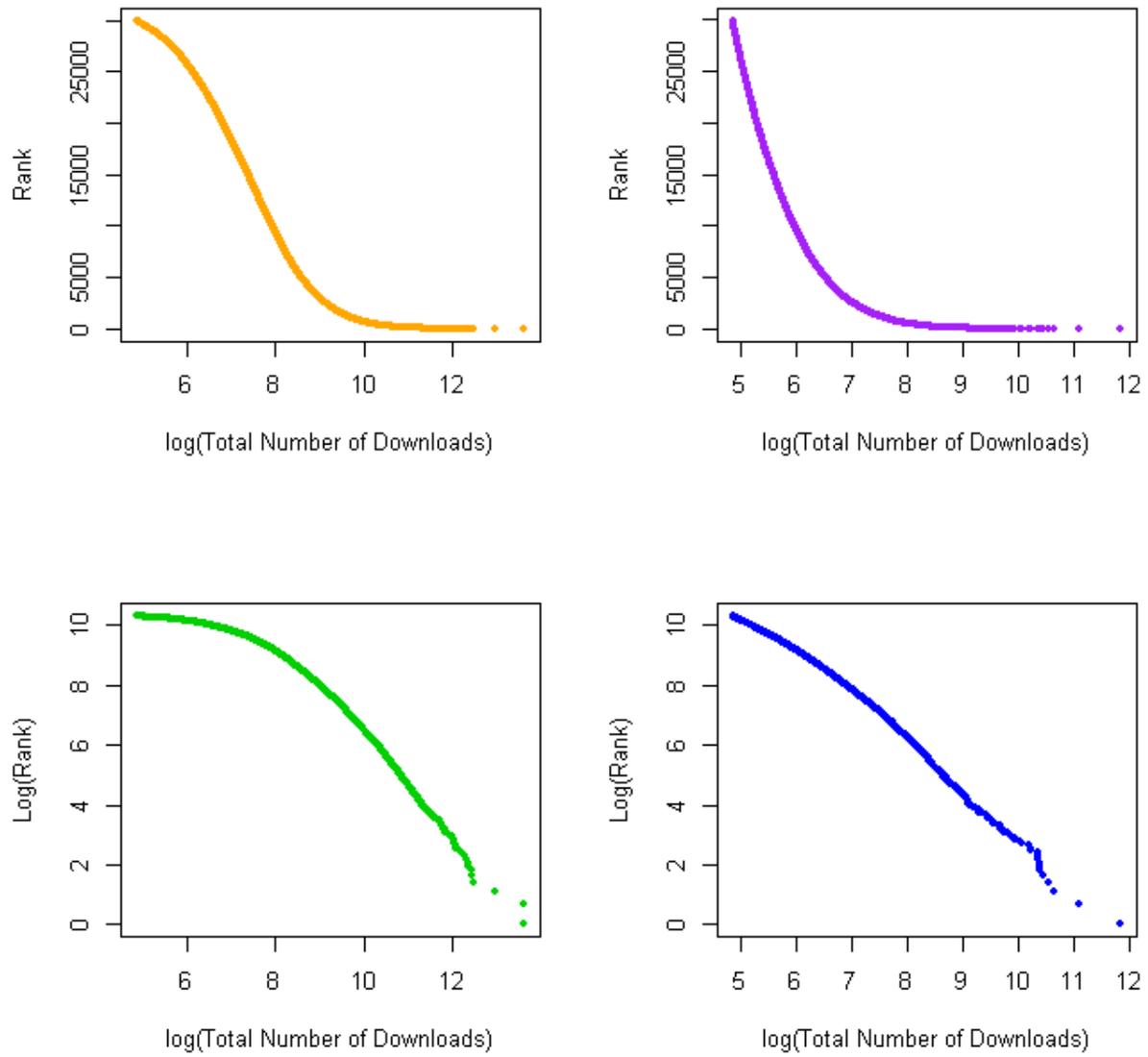

**Figure 5**. Downloads rank $r$ and $\ln(r)$ v. $\ln(d_{tot})$. Left column: overall; right column: the last 12 months. See Subsection 2.1 for details.

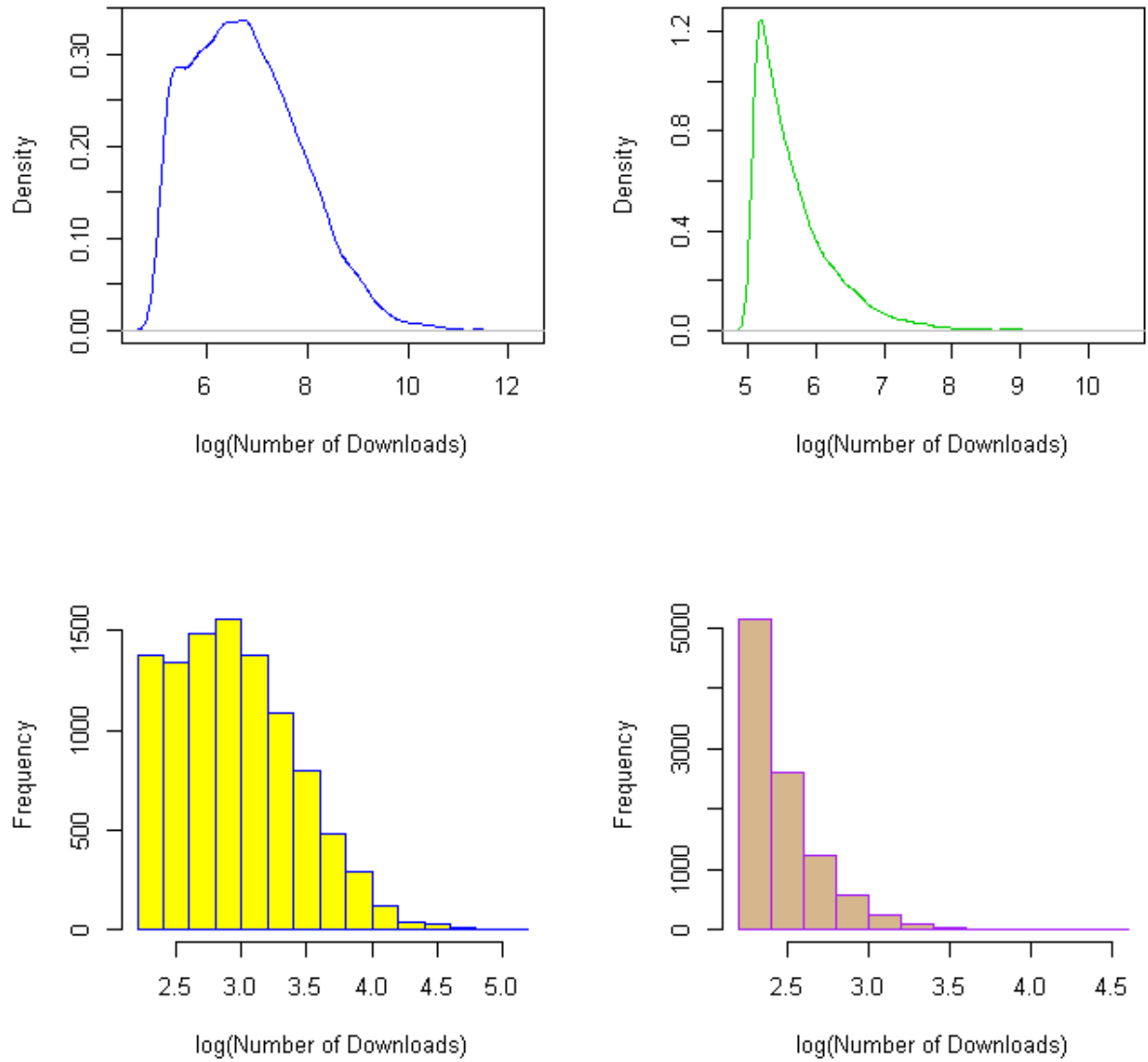

**Figure 6.** Cross-sectional (across all papers in the SSRN Top Papers data) density for $\ln(d_p)$ and histogram for $\log_{10}(d_p)$, where $d_p$ is the number of downloads for each paper. Left column: overall; right column: the last 12 months.

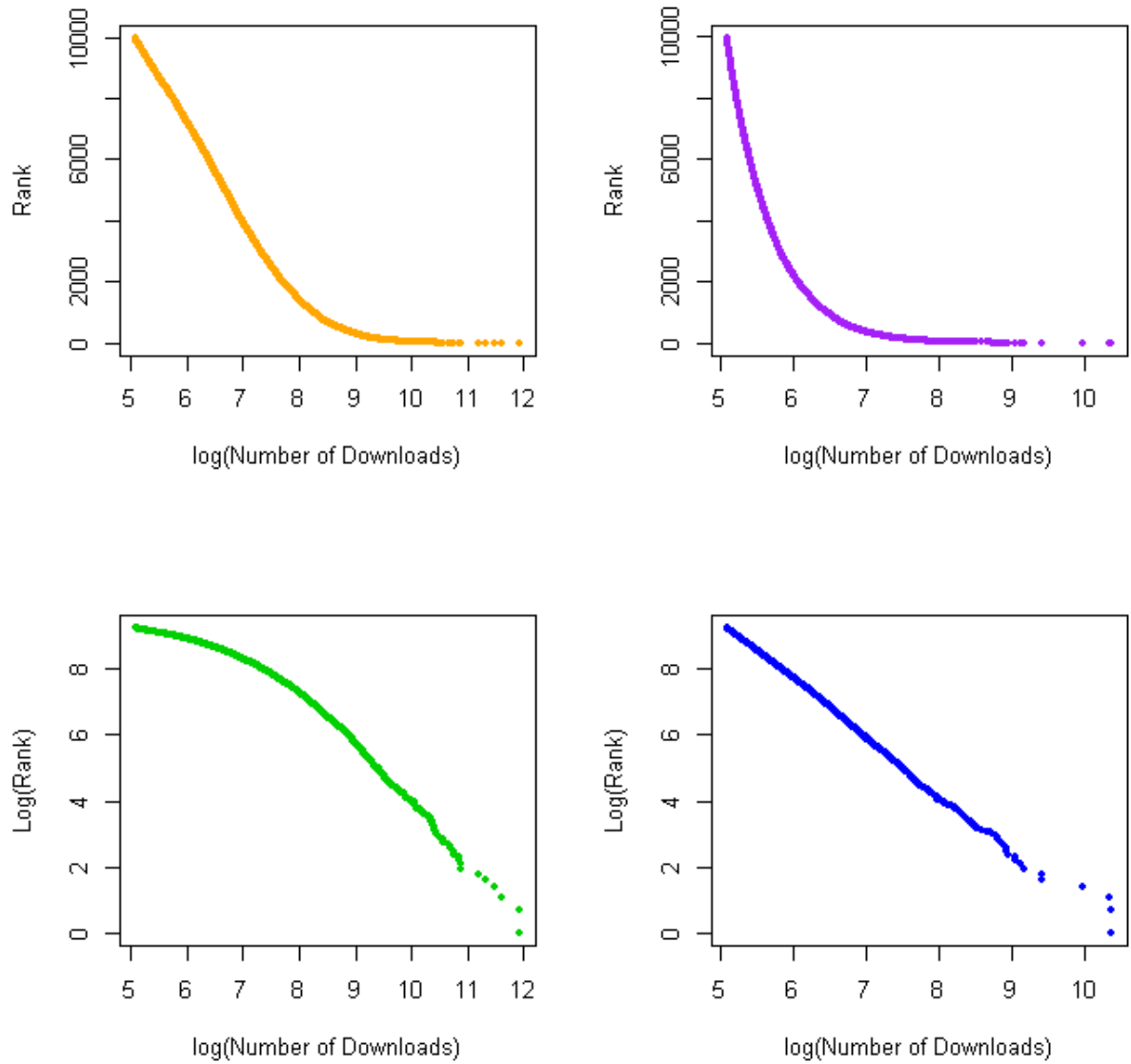

**Figure 7**. Downloads rank $r_p$ and $\ln(r_p)$ v. $\ln(d_p)$. Left column: overall; right column: the last 12 months. See Subsection 2.2 for details.

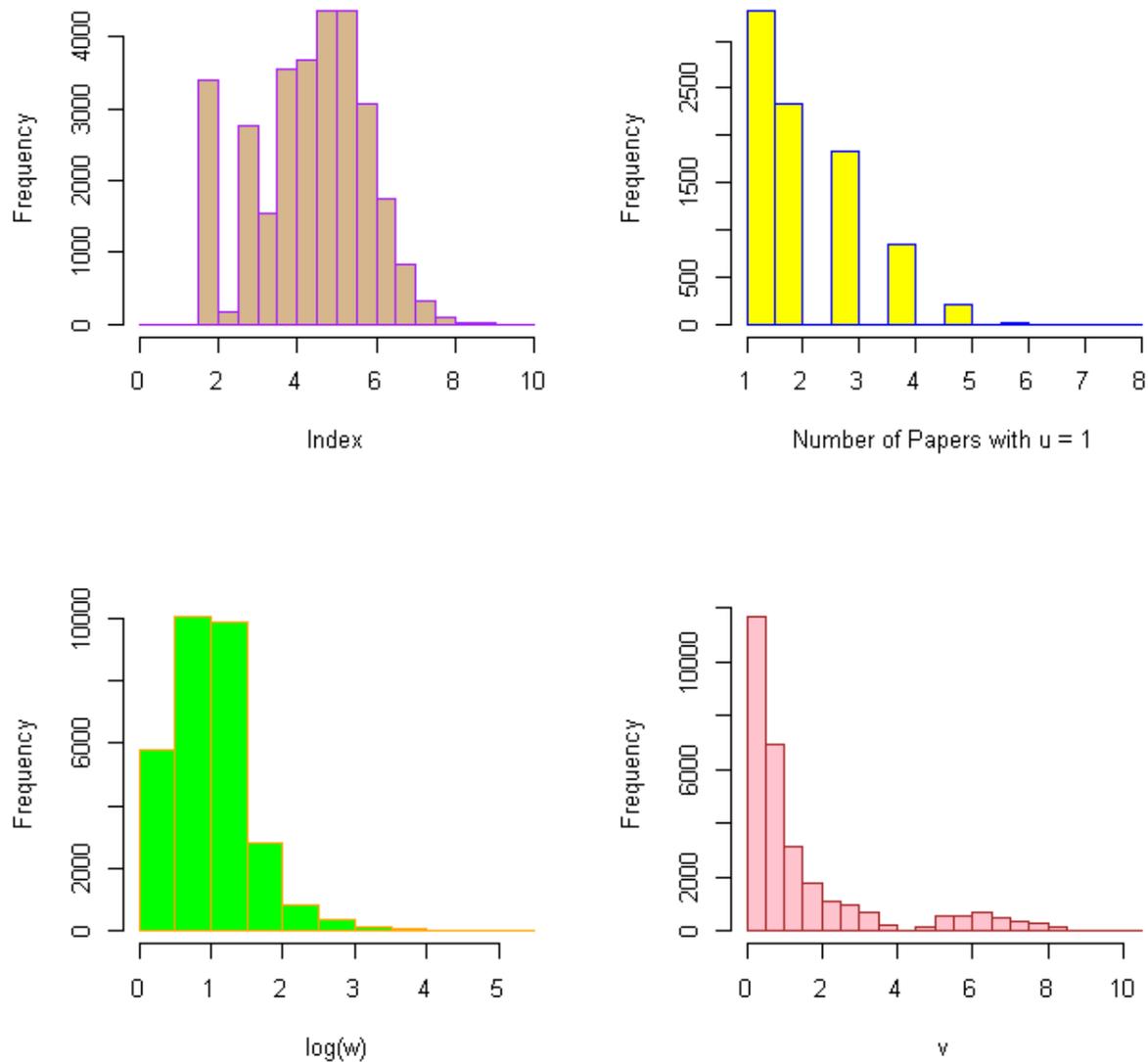

**Figure 8**. Upper-left: histogram of the $k_*$ index; upper-right: histogram of the number of papers $n_{tot}$ with $u = 1$ (where $u = k / n_{tot}$); lower-left: histogram of $\ln(w)$ (where $w$ is defined in Eq. (7)); lower-right: histogram of $v = \ln(d_{tot} / n_{tot}) / n_{tot}$. See Section 3 for details.

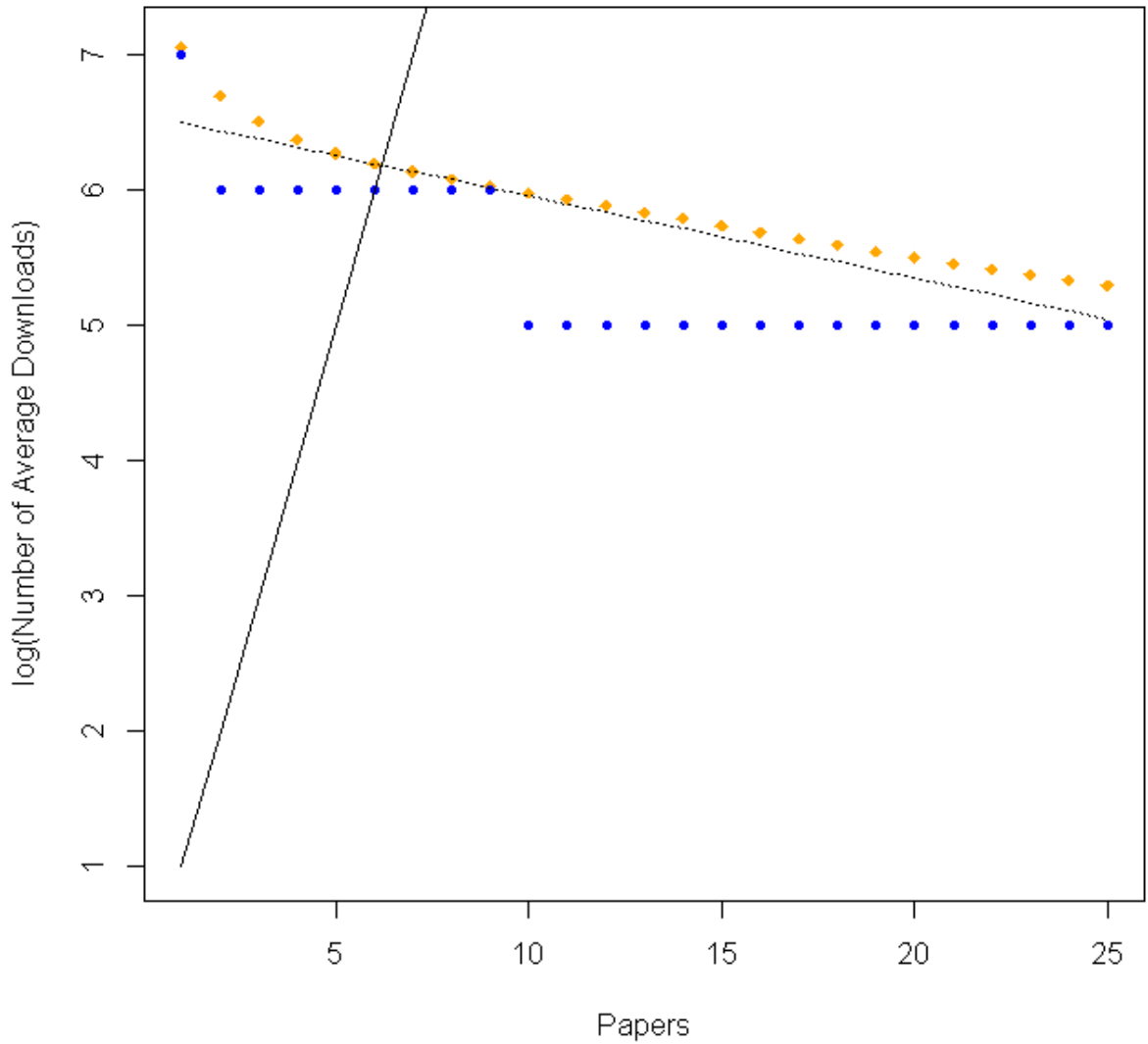

**Figure 9.** This figure illustrates the computation of the $\kappa$ and $\kappa_*$ indexes for the same randomly chosen author as in Figure 1. The sloping diamonds correspond to $\ln(f_i)$, where $f_i$ is the average number of downloads for the first $i$ papers (the papers are ordered decreasingly with the numbers of downloads $d_i$). The horizontal circles correspond to $t_i = \lfloor \ln(f_i) \rfloor$ in Eq. (9). The solid straight line has slope 1 and its intersection with the horizontal lines of circles gives the value of $\kappa = 6$. The dotted straight line with the negative slope goes through the points $(\kappa, \ln(f_\kappa))$ and $(\kappa + 1, \ln(f_{\kappa+1}))$ and its intersection with the solid straight line determines $\kappa_*$. In this example we have $\kappa_* = 6.182$ and $f_* = 484$.

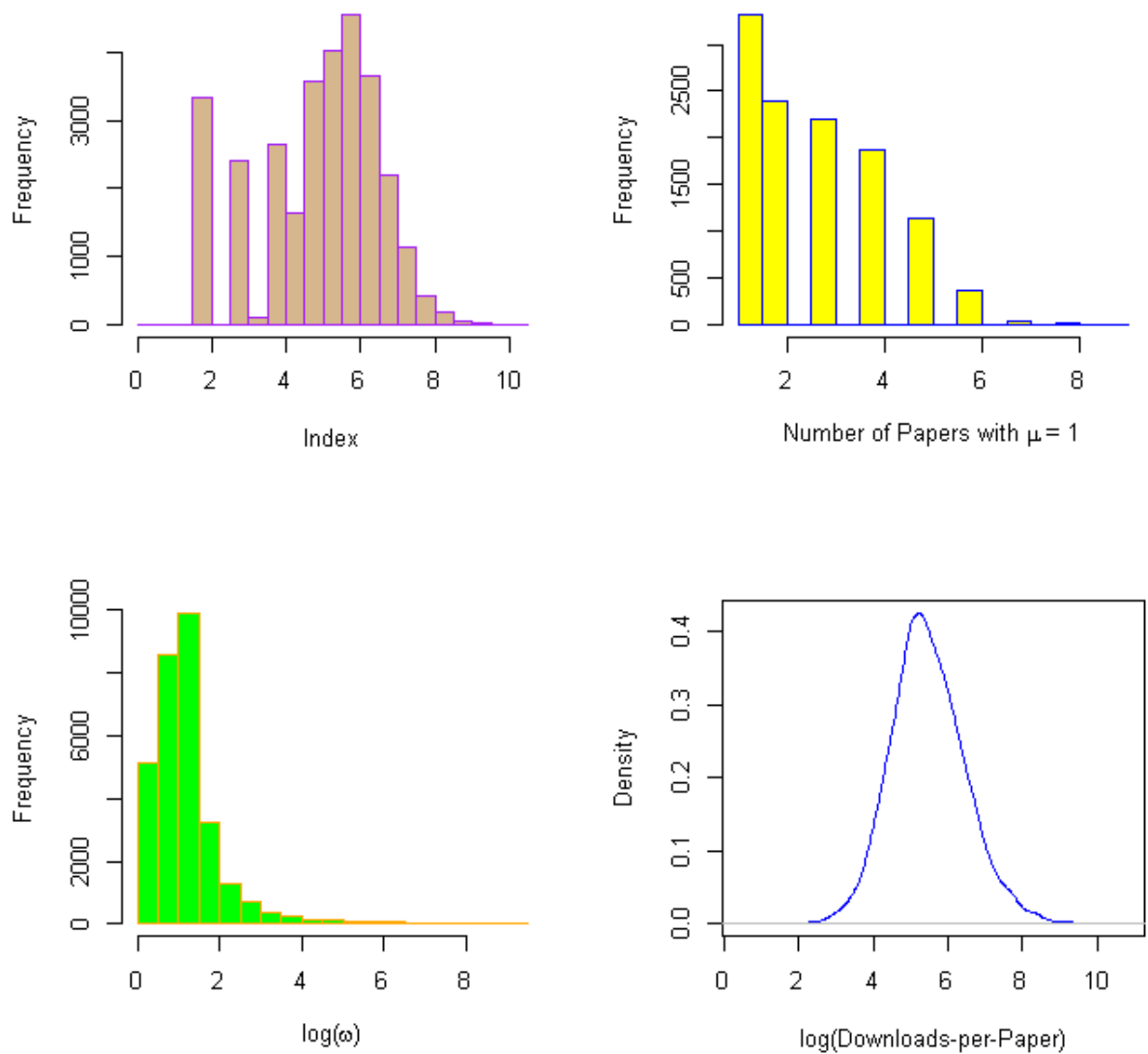

**Figure 10**. Upper-left: histogram of the $\kappa_*$ index; upper-right: histogram of the number of papers $n_{tot}$ with $\mu = 1$ (where $\mu = \kappa / n_{tot}$); lower-left: histogram of $\ln(\omega)$ (where $\omega$ is defined in Subsection 3.1); lower-right: density of $\ln(d_{tot} / n_{tot})$, where $n_{tot}$ excludes all papers with empty $d_i$ fields (so the latter do not alter the density curve shape, cf. Figure 3). See Subsection 3.1 for details.

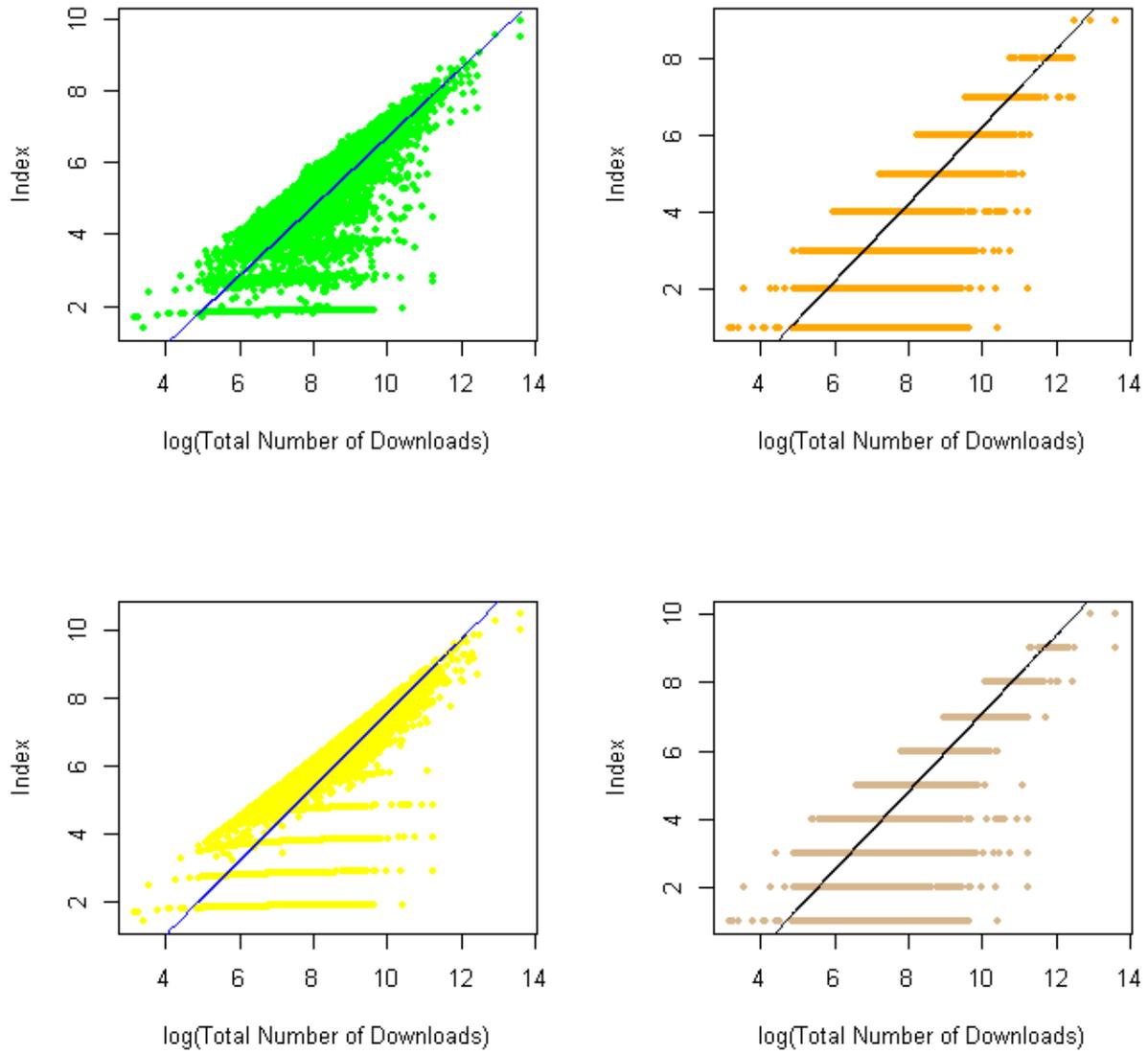

**Figure 11**. Upper-left: Index = $k_*$; upper-right: Index = $k$; lower-left: Index = $\kappa_*$; lower-right: Index = $\kappa$. Straight lines correspond to linear fits into the data (see Tables 12-15).

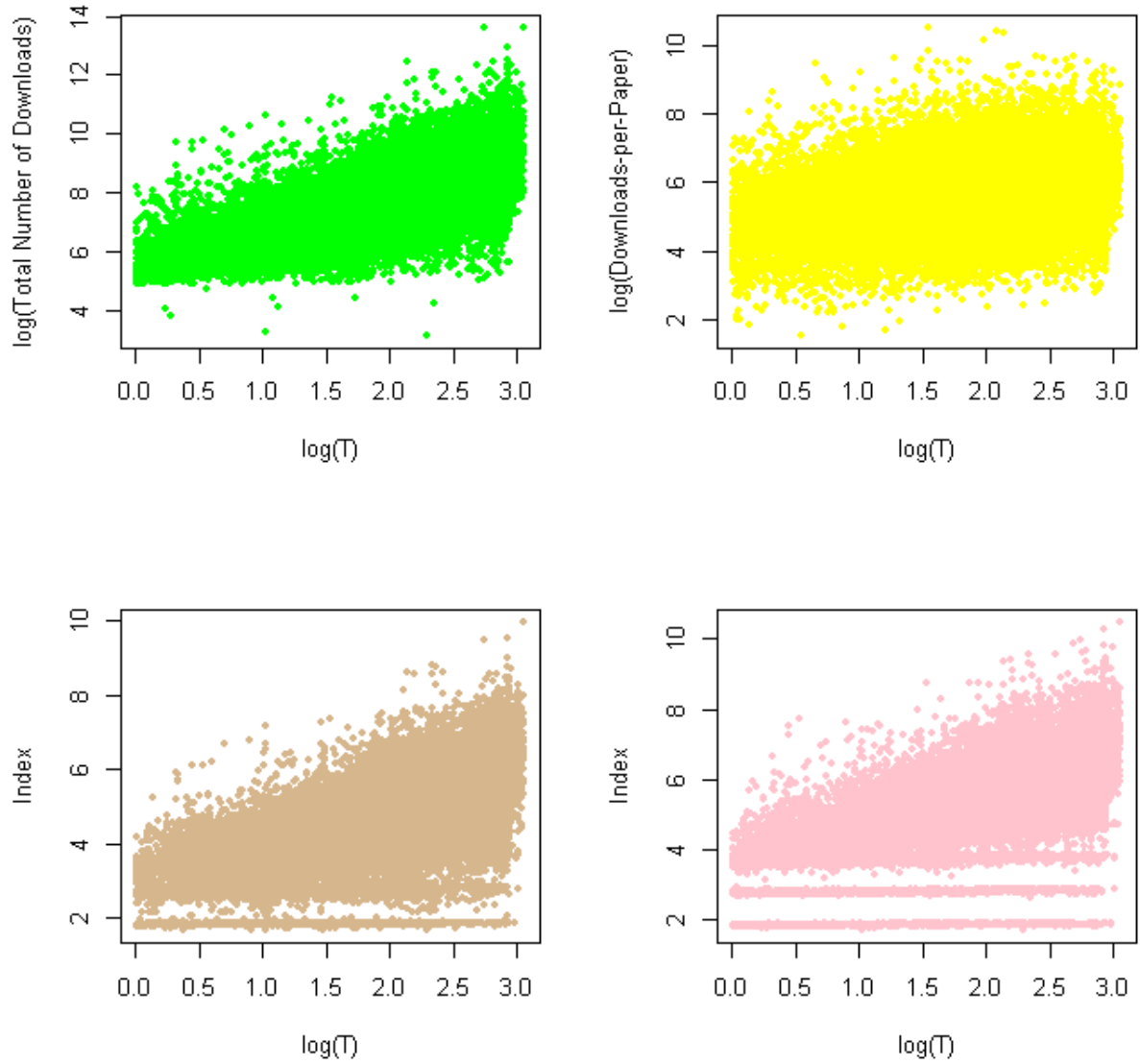

**Figure 12**. Upper-left: $\ln(d_{tot})$ v. $\ln(T)$; upper-right: $\ln(d_{tot} / n_{tot})$ v. $\ln(T)$; lower-left: $k_*$ v. $\ln(T)$; lower-right: $\kappa_*$ v. $\ln(T)$. Here $T$ is the time in (years) from the date of the author's first posting of a paper ("Posted:" field) on SSRN until August 17, 2015. Also see Table 20.